\documentclass[journal]{IEEEtran}

  \usepackage{amssymb}
  \usepackage{latexsym}
  \usepackage{amsfonts}
  \usepackage{amsthm}
  \usepackage{amsmath}
  \usepackage{graphics}
  \usepackage{setspace}
  \usepackage{graphicx}
  \usepackage{epstopdf}
  \usepackage{color}
  \usepackage{float}
  \usepackage{wrapfig}
  \usepackage{color}
  \usepackage{rotating}
  \usepackage{array} 
  \usepackage{dblfloatfix}
  \usepackage[hyphens]{url}
  \usepackage[table]{xcolor}
  \usepackage{multirow}

  \usepackage{graphics}
  \usepackage{setspace}
  \usepackage{algorithm}
  \usepackage{algorithmic}
  \usepackage{graphicx}
  \usepackage{epstopdf}  
  \usepackage{color}
  \usepackage{float}
  \usepackage{wrapfig}
  \usepackage{color}
  
  \usepackage{graphicx}
  \usepackage{caption}
  \usepackage{subcaption}
  \usepackage{setspace}
  
  \definecolor{blue}{rgb}{0,0,0}

\usepackage{lipsum}
  \usepackage{amsmath}

\definecolor{litered}{RGB}{246,194,197}
\definecolor{liteyellow}{RGB}{250,227,199}
\definecolor{litegreen}{RGB}{206,235,183}

\newcommand{\platform}{C-MOSDEN }

\newcommand{\crowdsensing}{\textit{crowdsensing }}

\ifCLASSINFOpdf
\else
\fi
\begin{document}
%
\title{Energy Efficient Location and Activity-aware On-Demand Mobile Distributed Sensing Platform  for Sensing as a Service in IoT Clouds}
%
%
%

\author{\IEEEauthorblockN{Charith Perera~\IEEEmembership{ Member,~IEEE}, Dumidu Talagala~\IEEEmembership{Member,~IEEE}, Chi Harold Liu~\IEEEmembership{Senior Member,~IEEE}, Julio C. Estrella~\IEEEmembership{Member,~IEEE} }





\thanks{C.~Perera is with the
Department of Computing, Faculty of Maths, Computing and Technology, The
Open University, Walton Hall, Milton Keynes, MK7 6AA, United Kingdom
(e-mail: charith.perera@open.ac.uk)}
\thanks{D. S. Talagala is with the Centre for Vision, Speech, and Signal Processing, University of Surrey, Guildford, Surrey GU2 7XH, U.K. (e-mail: d.talagala@surrey.ac.uk).}
\thanks{J. C Estrella is with Institute of  Mathematics and Computer Science (ICMC), University of  São Paulo, Brazil (email: jcezar@icmc.usp.br )}

\thanks{C. H. Liu is with Beijing Institute of Technology, China. (e-mail: chiliu@bit.edu.cn)}

\thanks{Manuscript received xxx xx, xxxx; revised xxx xx, xxxx.}}

%
%

\markboth{IEEE TRANSACTIONS ON COMPUTATIONAL SOCIAL SYSTEMS, VOL. xx, NO. xx, xxxxxxx xxxx}%
{Shell \MakeLowercase{\textit{et al.}}: Bare Demo of IEEEtran.cls for Journals}
%



\maketitle

\begin{abstract}

The Internet of Things (IoT) envisions billions of sensors deployed around us and connected to the Internet, where the mobile crowd sensing technologies are widely used to collect data in different contexts of the IoT paradigm. Due to the popularity of Big Data technologies, processing and storing large volumes of data has become easier than ever. However,  large scale data management tasks still require significant amounts of resources that can be expensive regardless of  whether they are purchased or rented (e.g. pay-as-you-go infrastructure). Further, not everyone is interested in such large scale data collection and analysis. More importantly, not everyone has the financial and computational resources to deal with such large volumes of data. Therefore, a timely need exists for a cloud-integrated mobile crowd sensing  platform that is capable of capturing sensors data, on-demand, based  on conditions enforced by the data consumers. In this paper, we propose a context-aware, specifically, location and activity-aware mobile sensing platform called C-MOSDEN (\textit{Context-aware Mobile Sensor Data ENgine}) for the IoT domain. We evaluated the proposed platform using three real-world scenarios that highlight the importance of \textit{selective sensing}. The computational effectiveness and efficiency of the proposed platform are investigated and is used to highlight the advantages of context-aware selective sensing.

\end{abstract}


%
\IEEEpeerreviewmaketitle


\section{Introduction}
\label{sec:Introduction}
\IEEEPARstart{T}{he} Internet of Things (IoT) \cite{myIoT} has become popular over the past decade. As part of the IoT infrastructure, sensors are expected to be deployed all around us, from everyday objects we use, to public infrastructure such as bridges and roads \cite{H1, H2}. As the price of sensors diminish rapidly, we can soon expect to see very large numbers of objects comprising of sensors and actuators. In addition, the modern technology-savvy world is already full of devices comprising of sensors, actuators, and data processors. The concentration of computational resources will enable the sensing, capturing, collection and processing of real time data from billions of connected devices , and can be envisaged to serve   many different applications including environmental monitoring, industrial applications, business and human-centric pervasive applications \cite{ZMP003}.

\textit{The Internet of Things allows people and things to be connected any time, any place, with anything and anyone, ideally using any path/network and any service} \cite{P029}. IoT is expected to generate large volumes of sensors data \cite{ZMP003}. Due to the latest innovations in the computer hardware sector and the reduction in  hardware costs, large scale data processing is becoming increasingly economical. Specially, with the popularity of utility-based cloud computing \cite{TCSS1} that offers computational resources in a '\textit{pay as you-go}' model, the tendency to collect a large amount of data has been increasing over the last few years. In 2010, the total amount of data on earth exceeded one zettabyte (ZB). By the end of 2011, the number grew up to 1.8 ZB \cite{ZMP003}. Further, it is expected that this number will reach 35 ZB in 2020. It is therefore apparent that sensor data has significant value if we can collect and extract insights from them.

Along with the IoT concepts, business models such as sensing as a service  has also generated significant interest \cite{ZMP008}.  The sensing as a service model  envisions a marketplace where sensor data is traded in an open and transparent manner with interested consumers. Sensing as a service can therefore  be seen as a platform where data owners can sell data to interested sensor date consumers in  'pay as you-go' fashion. On the one hand,  such a model stimulates the growth of sensor deployments. On the other hand, it reduces the cost of sensor data acquisition  due to its shared nature (i.e. sense once, sell to many). In addition, the sensing as a service model will also share the common IoT infrastructure to collect, process, and store data. In contrast, crowd sensing technologies have been widely used to collect sensor data in IoT paradigm. In community sensing, also referred to as group sensing \cite{P217} and mobile \crowdsensing \cite{crowdsensing}, the focus has been on monitoring of large-scale phenomena that cannot be measured using information from a single individual. The purpose here is to collect information from a large group of people in order to analyse and use that information for the benefit of the group  as a whole.

In the discussion so far, we  briefly introduced the IoT, sensing as a service model, and the Big Data in the IoT paradigm. In this paper, we define non-selective sensing as the process of collecting sensors data from all possible sensors available, all the time without any filtering. While we acknowledge the importance and  value of collecting large volumes of sensors data, a number of drawbacks of non-selective sensor data collection  exist.  Despite the fact that non-selective data collection could generate more value in  the long term (e.g. due to discovery of knowledge that were not intended during the time of data collection), it definitely  creates a problem (or difficulties) in the short term. The main issue in non-selective data collection is cost. Moreover, the processing and storing of data lead to more costs directly associated to the computational resource requirements (e.g. CPU, memory, storage space). Further, processing more data requires more time which creates the problem of not being able to extract knowledge from the collected data on time. Crucially, another issues is energy consumption. Sensors are typically resource constrained devices with limited access to energy. Non-selective sensing therefore leads to significant energy consumption and faster battery drain which create additional challenges related to the IoT infrastructure maintenance. Another challenge is  network communication. Large-scale data transfers over the network without any kind of filtering leads to the continuous use of the communication radios continuously. This also leads to faster battery drain in addition to the heavy network traffic generated in the IoT infrastructure. Thus, energy is a critical factor, especially in the crowd sensing domain, where humans  are involved in maintaining the sensing infrastructure. Therefore, we believe that on-demand selective sensing (i.e. perform sensing only under certain conditions) enables to avoid all the issues discussed above. To this end, we propose a scalable energy efficient data analytics platform for on-demand distributed mobile crowd sensing called \platform\footnote{It is also important to note that \platform is closely integrated into the GSN cloud middleware \cite{P022}.}.  

The rest of this paper is organised as follows. In Section \ref{sec:Problem}, we define the problem domain in details. The functional requirements of the proposed solutions is presented in Section \ref{sec:Functional}. The proposed mobile crowd sensing platform is explained in detail in Section \ref{sec:Solution}. The cost models and the advantages of using the proposed platform is discussed in Section \ref{sec:Model}. Section \ref{sec:Implementation} discusses the implementation details. Experimentation and evaluation details are presented in Section \ref{sec:Evaluation}.  Related works are discussed in Section \ref{sec:Related_Work}. Finally, Section \ref{sec:Conclusions and Future Work} concludes the paper.

\section{Problem Definition and Motivation}
\label{sec:Problem}

In the earlier section we briefly introduced our problem domain. In this section, we explain the problem we address in this paper in detail.

The mobile crowd sensing technologies are widely used to collect data in different contexts in the IoT paradigm. Due to popularity of Big Data technologies, processing and storing large volumes of data has become easier than ever. However, still such large scale data management tasks are economically costly. For example, Microsoft Azure\footnote{http://www.windowsazure.com/en-us/pricing/details/virtual-machines/} cloud computing platform charges  541 USD/month for 8 cores and 14GB RAM. Google\footnote{https://cloud.google.com/products/app-engine/} cloud services  pricing is similar. Not everyone is interested in such large volumes of data collection and analysis. Further, not everyone has the financial and computational resources to deal with large volumes of data. Therefore, there is a real need for a mobile crowd sensing  platform that is capable of capturing sensor data  on-demand based on user requests and the conditions imposed by the data consumers.

\begin{figure}[t]
 \centering
 \includegraphics[scale=0.78]{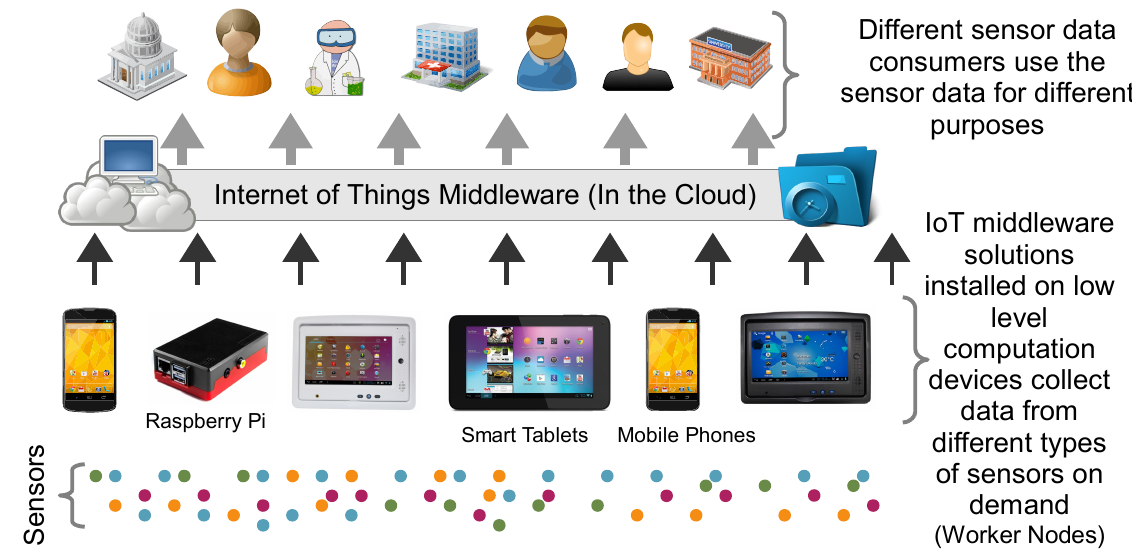}
 \caption{The proposed platform can be installed on both mobile and static resource constrained devices. The platform provides easy ways to connect sensors. Each of this platform instances act as worker nodes and able to carry out sensing tasks as directed by the cloud-based IoT middleware.}
 \label{Figure:Sensing_as_a_service}	
\end{figure}

Sensing as a service model, as illustrated in Figure \ref{Figure:Sensing_as_a_service}, shows how cloud IoT middleware (e.g. GSN \cite{P022}) works hand-in-hand with multiple worker nodes (e.g. C-MOSDEN). We identify two fundamental components in this sensing as a service architecture: 1) the cloud platform which manages and supervises the overall sensing tasks, and 2) worker nodes that actually perform the sensing tasks as instructed by the cloud IoT platform. It is important to note that our objective is not to analyse the data and extract any knowledge. In this context, our objective here is to collect only the most important and relevant data so the interested data consumers can use the data to extract the knowledge in an efficient manner with minimum use of computational resources, energy, time and labour. Our proposed platforms is ideal to be installed on worker nodes. Further, IoT middleware platforms such as Global Sensor Network (GSN) \cite{P022} can be used as the cloud middleware. The cloud IoT middleware evaluates the availabilities of worker nodes and sends the requests to a specific number of selected worker nodes. More importantly, sensor data consumers may impose specific conditions on the data acquisition or transfer,  such as \textit{`sense only when  a certain activity occurs'}. The detailed functional requirements of this system are discussed in Section \ref{sec:Functional}.

\section{Functional Requirement Analysis}
\label{sec:Functional}

In this section, we discuss some of the major functional requirement of a worker node in an ideal on-demand mobile crowd sensing platform. Let us consider three different scenarios from three different domains: 1) environmental monitoring, 2) physical rehabilitation  and 3) health and well-being.

\begin{figure}[b]
 \centering
 \includegraphics[scale=0.76]{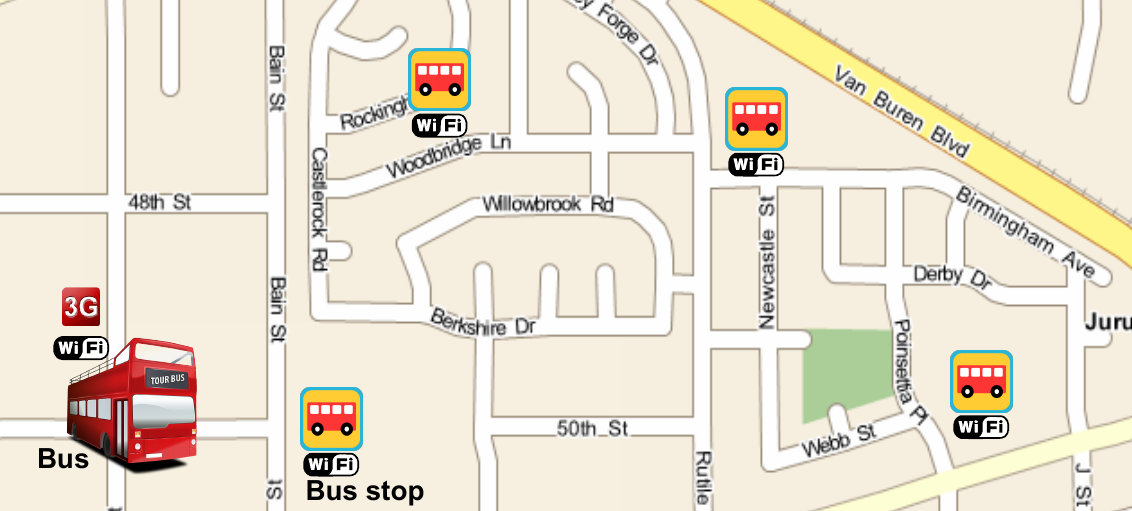}
 \caption{Usecase Scenario 1: Sensors are deployed in buses in a Smart City environment and the data is expected to be collected based on context information and conditions provided by the data consumer.}
 \label{Figure:Usecase1}	
\end{figure}

\textbf{Scenario 1 (Environmental Monitoring):} John, a researcher at the Department of the Environment, is interested in measuring and monitoring the air pollution in cities. John's team has deployed sensor kits in buses. Each of these sensors kits consists of multiple sensors and a communication device with both WiFi and 3G capabilities. John's team has developed an application that processes data collected by these sensor kits. This application consists of a number of different algorithms that analyse and visualise air pollution in the city. However, according to the way that the  algorithms are written, John only needs to collect data when the buses are moving. Sensor data captured while the bus is stopped at a bus stop, or in traffic does not add any value. Therefore, John would like to collect sensor data only when the bus is moving. Further, John does not need real-time data in most of the occasions. Therefore,  it is sufficient to push the  sensor data to the cloud when the bus researches a bus stop. The communication devices fitted in the bus will connect to the bus stop's WiFi and push the data collected since the last bus stop, as illustrated in Figure \ref{Figure:Usecase1}. However, John is also interested to receive sensor data in real-time when raining. Therefore, when raining, the communication devices need to use 3G to upload the sensor data to the cloud. However, they still need to adhere to the first rule  that says \textit{`sense only when moving'}.

\begin{figure}[h]
 \centering
 \includegraphics[scale=0.78]{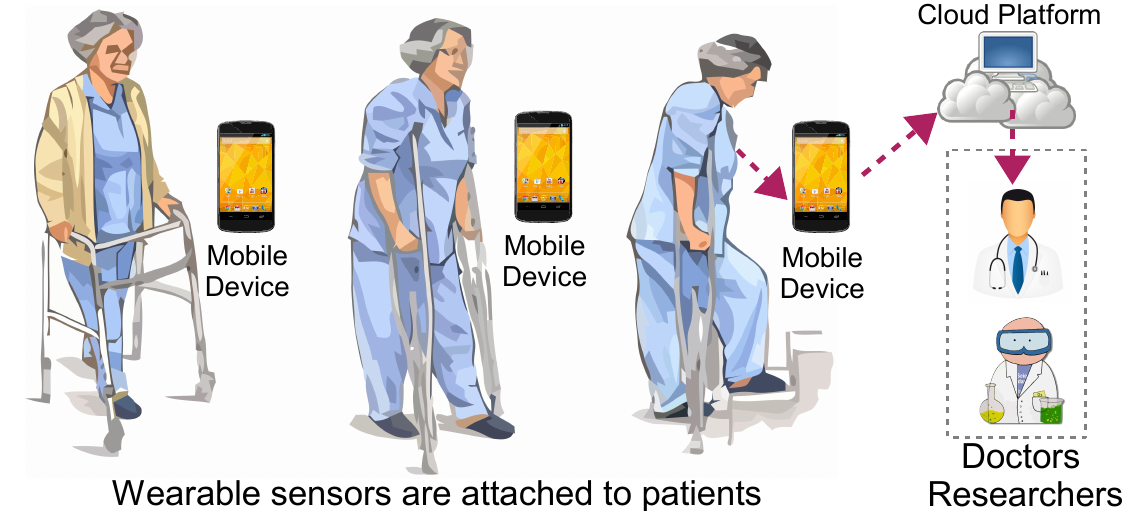}
 \caption{Usecase Scenario 2: Wearable sensors are attached to patients body. Doctors and researchers are expected to collect data from the sensors based on context information.}
 \label{Figure:Usecase2}	
\end{figure}

\textbf{Scenario 2 (Rehabilitation):} Robert is a researcher who oversees a number of a rehabilitation facilities around the country where patients with physical disabilities are treated and rehabilitated. Robert is interested in collecting sensor data from sensors worn by patients while they engage in certain activities. Robert has  developed an application that requires data collected from wearable sensors \cite{bodyArea} only when patients are walking and climbing stairs (as part of the exercise programs recommended by doctors). Wearable sensor kits push data to the patient's smartphone. Each smartphone pushes data to the cloud when it get access to the Internet as illustrated in Figure \ref{Figure:Usecase2}. It is important to note that Robert is only interested in data collection  when the patients are performing certain activities. The blind collection of data during other times may impact negatively in the analysis done by the application Robert has developed. Such data also wastes time and resources in the event that Robert has to filter the data he wants from large volume of irrelevant data.

\begin{figure}[h]
 \centering
 \includegraphics[scale=0.78]{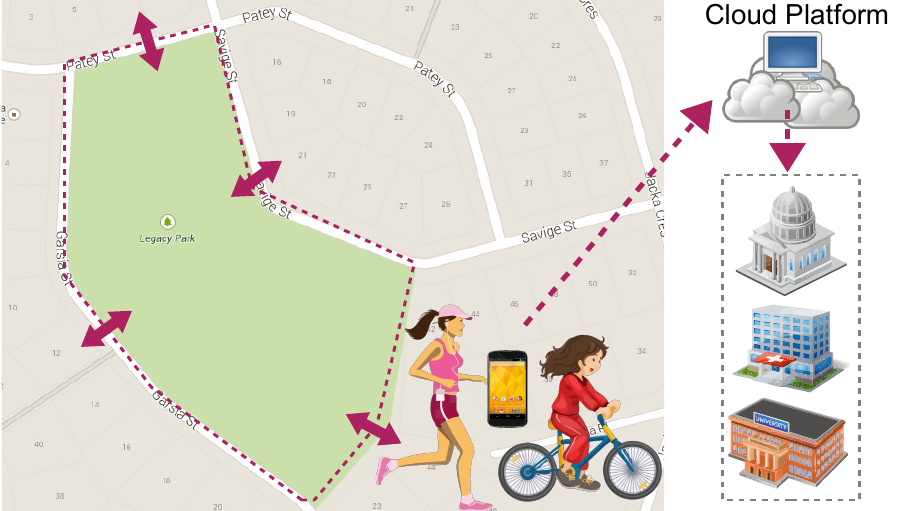}
 \caption{Usecase Scenario 3: Wearable sensors can be used to monitor movements of the  general public who use public spaces such as parks for exercising and recreational activities so the authorities can plan further development and upgrades of the infrastructure.}
 \label{Figure:Usecase3}	
\end{figure}

\textbf{Scenario 3 (Health and Well-being):} Michael is working for the Department of public health and well-being. He has been asked to develop a plan to improve the public health  in cities by improving the infrastructure that supports  exercise and recreational activities (e.g. parks and the paths that supports jogging, cycling, and venues for bar exercise). Michael developed a wearable lite-weight sensor kit that can monitor a variety of different parameters such as air quality, sound, movement. Further, Michael has recruited volunteers who are willing to wear those sensor kit when  exercising. The sensor kit collects data and pushes it to the volunteer's smartphone. A smart phone application push data to the cloud once it gets connected to the Internet. However, Michael  only needs to collect data when a volunteer enters  the park areas as illustrated in Figure \ref{Figure:Usecase3}. Further, Michael only needs to perform sensing only when the volunteers are moving (e.g. walking, running, cycling). Michael has notices that there  is a large amount of people coming to the park during the weekend. In order to reduce the burden to the volunteers, Michael only needs to collect data from a maximum of 30 sensors  kits (i.e. volunteers) despite the actual number of volunteers visiting the park in weekends.

As you may have already noticed, each of these scenarios required sensing to be performed under different conditions. Due to limited resources typically, researchers try to limit the data they collect such that the processing can be done with limited resources. Further, due to the ways that analysis tools are programmed, the data they accept may differs. They tend to perform well and produce accurate results when filtered, relevant data are provided. The most widely needed filtering conditions in mobile crowd sensing platforms are location-awareness (i.e. spatial),  activity-awareness, time-awareness (i.e. temporal), and energy-awareness \cite{SeCoMan}. Therefore, we developed our proposed platform, C-MOSDEN, to facilitate all of these conditions. Further, it is evident that we require cyber physical systems \cite{Khaitan}, which have both physical sensing components and cloud based data analysis software modules, to accommodate the scenarios we described earlier in this section.

\section{Proposed Sensing Platform}
\label{sec:Solution}

\begin{figure}[h]
 \centering
 \vspace{-0.23cm}
 \includegraphics[scale=0.40]{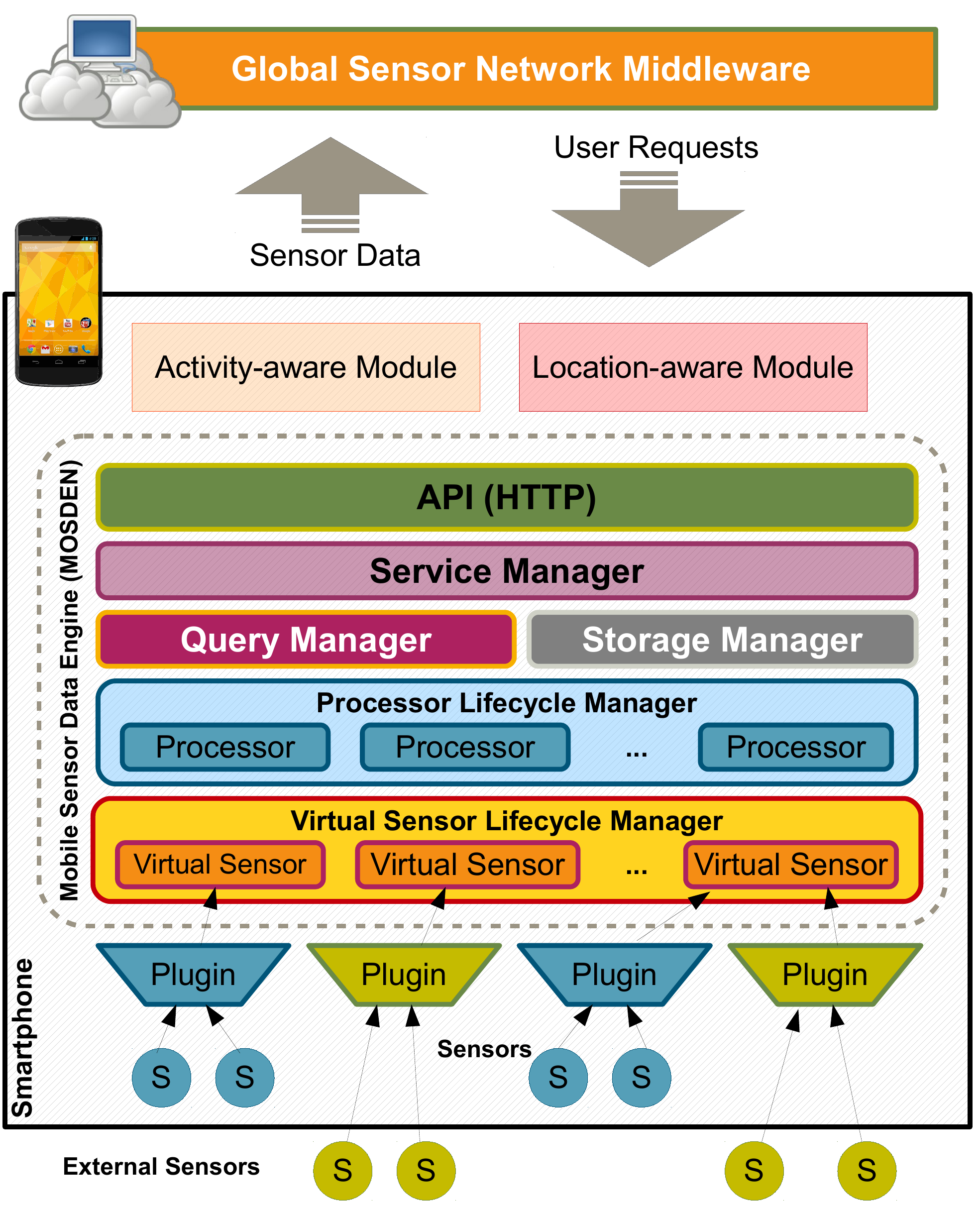}
 \caption{The Proposed C-MOSDEN Platform}
 \label{Figure:Architecture}	
\vspace{-0.53cm}	
\end{figure}

In order to address the challenges discussed in the previous section, we propose a novel mobile sensing platform called \platform. It consists of three components: 1) Context-aware data streaming engine called Mobile Sensor Data Engine (C-MOSDEN). C-MOSDEN platform is based on our previous platform MOSDEN \cite{ZMC008}. MOSDEN is an IoT middleware for resource constrained devices, that allows to collect and process sensor data without programming efforts. Sensing as a service model \cite{ZMP008} is a first class citizen in MOSDEN design.  MOSDEN is a client-side tool that can be installed in any android device such as smart phones and DragonBoard 410c. However, MOSDEN does not facilitate any context-aware functionalities. Instead, MOSDEN is a mechanism to filter data streams based on users defined queries. In C-MOSDEN,  new  querying capabilities are introduced to support context-aware functionalities., 2) the activity-aware module, and 3) a location-aware module. The complete architecture of the proposed platform, C-MOSDEN, is presented in Figure \ref{Figure:Architecture}. First, we discuss the three main components  in the subsequent sections in brief. Then, we briefly introduce the IoT cloud middleware employed, called Global Sensor Network Middleware (GSN) \cite{P022},  which is the cloud-based companion platform of the proposed mobile sensing system. At the end, we explain how GSN and C-MOSDEN work together as a system to achieve a common objective.

\subsection{Context-aware Mobile Sensor Data Engine (C-MOSDEN)}
C-MOSDEN is a plug-in-based IoT middleware for mobile devices (e.g. mobile phones, tablets, Raspberry Pi like platforms) that allows to collect and process sensor data without programming efforts. C-MOSDEN is also a true zero programming middleware where users do not need to write program code or any other specifications using declarative languages. C-MOSDEN also supports both push and pull data streaming mechanisms as well as centralised and decentralised (e.g. peer-to-peer) data communication. Plugins can be installed separately to extend the capabilities of C-MOSDEN (e.g. provide support for difference types of sensors). This engine supports a number data processing and filtering capabilities such as comparison operators, average, etc. Additionally, it supports scalable and distributed sensing. C-MOSDEN can handle more than 100 user request at a given time. Performance evaluation details are presented in \cite{ZMC008, ZMP010}. In C-MOSDEN, mobile devices are configured using a human readable SQL-like query language as explained following sections.

\subsection{Activity-aware Module}
\label{sec:Activity}
The activity-aware module is capable of recognizing six different activities. The detectable activities are 1) moving in a vehicle, 2) cycling, 3) walking, 4) running, 5) still (not moving), 6) tilting (falling). Therefore, users can combine these activities to build different types of queries.

\subsection{Location-aware Module}

The location-aware module is capable of recognizing when the device moves into a certain area and moves away from a certain area. These locations are defined using latitude, longitude, and radius in meters. Location-awareness can also be combined with other sensing parameters as presented in Figure \ref{Figure:Architecture}.

\begin{figure}[h]
 \centering
 \includegraphics[scale=0.84]{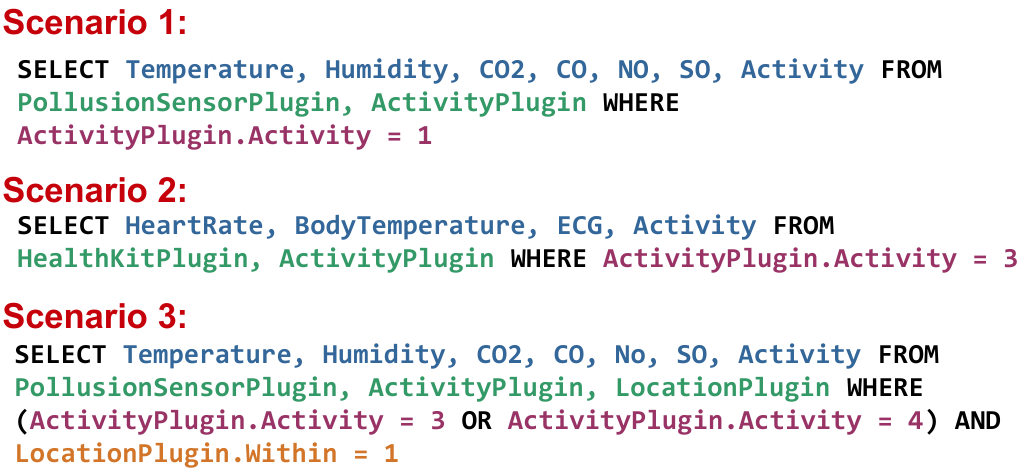}
 \caption{Sample queries related to the three scenarios presented earlier in the paper. Queries are slightly modified for demonstration and clarity purposes.}
 \label{Figure:Sample_Query_Snippits}	
\end{figure}

Let's consider three different queries built to support the three scenarios presented in Section \ref{sec:Functional}. Figure \ref{Figure:Sample_Query_Snippits} illustrates how to combine different sensing parameters including both activity and location awareness.

Mostly the sample queries are self descriptive. However, it is important to note that activities are represented by numbers as explained in Section \ref{sec:Activity}. Further, 'within' location is defined as a boolean value (true = 1 and false = 0). Sensor data retrieved by different plugins\cite{ZMC008} can be referred in the query. For example, in query 2, health data is retrieved though a plugin  called \textit{HealthKit}. This means C-MOSDEN is retrieving data from a external IoT health product \cite{ZMJ007}.

\subsection{Global Sensor Network Middleware (GSN)}
The Global Sensor Network (GSN) \cite{P022} is an IoT cloud platform aimed at providing flexible middleware to address the challenges of sensor data integration and distributed query processing. It is a generic data stream processing engine. GSN has gone beyond the traditional sensor network research efforts such as routing, data aggregation, and energy optimisation. The design of GSN is based on four basic principles: simplicity, adaptivity, scalability, and light-weight implementation. GSN middleware simplifies the procedure of connecting heterogeneous sensor devices to applications. Specifically, GSN provides the capability to integrate, discover, combine, query, and filter sensor data through a declarative XML-based language and enables zero-programming deployment and management. The GSN is based on a container based architecture. A detailed explanation is provided in \cite{P022}. The \textit{Virtual Sensor} is the key element in the GSN. A virtual sensor can be any kind of data producer, for example, a real sensor, a wireless camera, a desktop computer, a mobile phone, or any combination of virtual sensors. Typical, a virtual sensor can have multiple input data streams but have only one output data stream.  In this work, GSN plays the   role of cloud in the IoT platform. It provides the functionality of global scheduler that manages worker nodes. In the next section, we explain how GSN and C-MOSDEN work together as a system.

\subsection{System Work Flow}

\begin{figure}[t]
 \centering
 \vspace{-0.43cm}
 \includegraphics[scale=0.78]{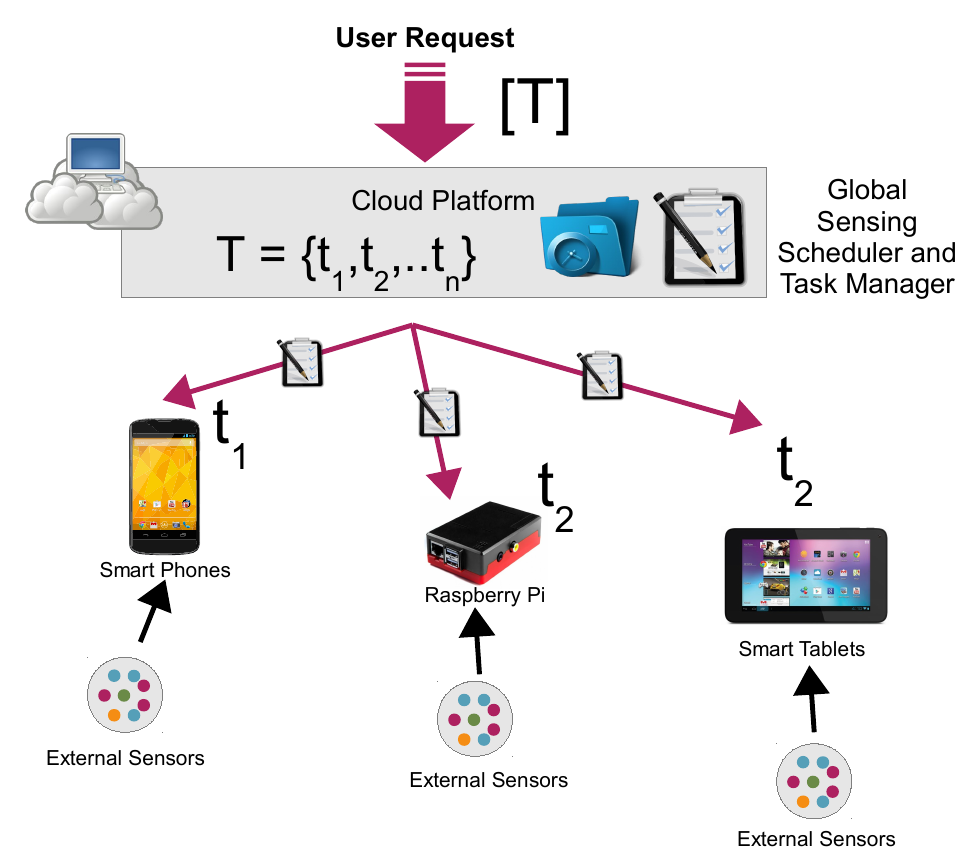}
 \caption{The IoT cloud platform is responsible for global scheduling sensing tasks. For example, GSN performs that task.  The IoT cloud platforms can send the specific sensing objectives to the clients, so the C-MOSDEN exactly sends back only the data that is requested}
 \label{Figure:Task_Scheduler}	
\end{figure}

As illustrated in Figure \ref{Figure:Task_Scheduler}, first sensor data consumers (e.g. city council, researcher, medical doctor) submit their requirement. Then, the IoT cloud platform analyses the consumer's problem and decides which sensors are to be used to collect relevant data \cite{ZMP009}. Then, the global task scheduler develops a strategic plan on how to delegate the tasks to multiple worker nodes (e.g. C-MOSDEN). Finally, global scheduler sends individual requests to a selected number of worker nodes (e.g. C-MOSDEN). These requests provide exact specifications and sensing objectives on  how, when, and where to collect data. Context-awareness allows to eliminates significant amount of uninterested data from communicating over limited network resources. The next section will present theoretical models for the resources saving of context-aware selective sensing. Experimental evaluation to justify the theoretical models are discussed in Section \ref{sec:Evaluation}.

\section{Cost Model for Efficient Sensing}
\label{sec:Model}

In this section, we develop cost models to evaluate the efficiency of C-MOSDEN from three different perspectives: 1) energy, 2) storage, and 3) network communication.

\subsection{Energy Consumption Modelling}
\label{sec:Energy}

The notations will be described  as we introduce them in the upcoming sections. As denoted in Equation \ref{equation:01}, the total energy consumption of a mobile sensing platform at a given point in time (i.e. $\Delta$) depends on two factors: 1) energy used for computational tasks (denoted by $E_{CPU}^{\Delta}$), and 2) energy used for data communication tasks (denoted by $E_{DCom}^{\Delta}$). It is also important to note that the data communication can also be divided into two parts: 1) ($E_{DCom^{S2D}}^{\Delta}$) data communication between sensors (S) and the local computational device (D)  (e.g.  between external sensors and the mobile phone \cite{ZMPB001}), and 2) ($E_{DCom^{D2C}}^{\Delta}$) data communication between the computational device (D) and the IoT cloud  middleware (C) as defined in Equation \ref{equation:01_2}. Further, we can define $E_{DCom}^{\Delta}$ based on the communication protocols as well, as presented by Equation \ref{equation:01_3} (e.g. 3G, WiFi, Bluetooth, ZigBee, Z-Wave, etc.). Typically, long range protocols consumes significantly more energy than short range protocols. However, there are some other energy costs as well (e.g. operating system related computational tasks, display, and so on.) where we denote them using a constant $K^{\Delta}$ in Equation \ref{equation:01}.

\begin{equation}
\label{equation:01}
E_{Total}^{\Delta} = E_{CPU}^{\Delta} + E_{DCom}^{\Delta}   + K^{\Delta}
\end{equation}

\begin{equation}
\label{equation:01_2}
E_{DCom}^{\Delta}  =  E_{DCom^{S2D}}^{\Delta} + E_{DCom^{D2C}}^{\Delta}
\end{equation}

\begin{equation}
\label{equation:01_3}
E_{DCom}^{\Delta}  =  E_{3G}^{\Delta} + E_{WiFi}^{\Delta} +  E_{BT}^{\Delta} + E_{ZigBee}^{\Delta} + E_{Z-Wave}^{\Delta}
\end{equation}

In Section \ref{sec:Functional}, we presented three different scenarios. Each of these scenarios had their own set of requirements and sensing objectives regarding how, what, and when to collect data. A scenario can be considered as an experiment that takes place within a certain period of time. We use $S^{\Theta}$ to denote a scenario. Equation \ref{equation:02} defines the total energy consumption by the scenario $S^{\Theta}$.

\begin{equation}
\label{equation:02}
E_{Total}^{S^{\Theta}} = E_{CPU}^{S^{\Theta}} + E_{DCom}^{S^{\Theta}}  +  K^{S^{\Theta}}
\end{equation}

In Equation \ref{equation:03}, we introduce $S^{\Psi}$ instead of $S^{\Theta}$. In the scenario defined in  Equation \ref{equation:02}, sensor data is collected using non-context-aware fashion. This means that the mobile sensing platform has been configured to collect data during the total time of the experiment (no activity-aware or location-aware capabilities have been used). In contrast, Equation \ref{equation:03} defines the total energy consumption when context-aware capabilites are activated. As  mentioned earlier, the context-aware capabilites are provided by the C-MOSDEN platforms at some cost. For example, in order to provide activity-aware and location-aware services, C-MOSDEN needs to perform some additional computations. Such additional computations need to be added to the total energy consumption equation. We use $E_{\Omega}^{S^{\Psi}}$ to denote such overhead computational costs.

\begin{equation}
\label{equation:03}
E_{Total}^{S^{\Psi}} = E_{CPU}^{S^{\Psi}} + E_{DCom}^{S^{\Psi}} +  K^{S^{\Psi}} + E_{\Omega}^{S^{\Psi}}
\end{equation}

The total energy consumption by the CPU when context-aware capabilities are not in use (i.e. scenario $S^{\Theta}$) denoted by Equation \ref{equation:04}. $PT_{CPU}^{S^{\Theta}}$ denotes the CPU processing time of the scenario $S^{\Theta}$. $E_{CPU}^{\Delta}$ denotes the energy cost at a given time. Therefore, total energy consumption by the CPU during a the scenario $S^{\Theta}$ is denoted by $E_{CPU}^{S^{\Theta}}$.

\begin{equation}
\label{equation:04}
E_{CPU}^{S^{\Theta}} =  E_{CPU}^{\Delta} \times PT_{CPU}^{S^{\Theta}}
\end{equation}

Similarly, Equation \ref{equation:05}  denotes the total energy cost for data communication. At a given point of time  data communication energy cost is $E_{DCom}^{\Delta}$. Total data transmission time is denoted by $TT_{DCom}^{S^{\Theta}}$.

\begin{equation}
\label{equation:05}
E_{DCom}^{S^{\Theta}} =  E_{DCom}^{\Delta} \times TT_{DCom}^{S^{\Theta}}
\end{equation}


It is important to note that in scenario $S^{\Theta}$, the  data communication is performed throughout the total duration (due to non-selective, non-context-aware sensing strategy).

For example, let us consider data communication related  energy consumption. $TT_{DCom}^{S^{\Theta}} $ is denoted by Equation \ref{equation:07}. The total duration of the scenario is denoted by $T_{TD}^{S^{\Theta}}$. The network communication frequency (i.e. how frequently the data needs to be sent to the cloud) is denoted by $T_{NCF}^{S^{\Theta}}$. Therefore, the number of time that the mobile device needs to push data to the cloud is denoted by $\frac{T_{TD}^{S^{\Theta}} }{T_{NCF}^{S^{\Theta}}  }$. $T_{DCom}^{\alpha }$ denotes the time it takes to push data to the cloud for one single round. It is important to  note that we mainly consider the data communication between the local computational device and the IoT cloud (i.e. $E_{DCom^{D2C}}^{\Delta}$) due to its significance over $E_{DCom^{S2D}}^{\Delta}$.

\begin{equation}
\label{equation:07}
TT_{DCom}^{S^{\Theta}}  = \frac{T_{TD}^{S^{\Theta}} }{T_{NCF}^{S^{\Theta}}  } \times T_{DCom}^{\alpha }
\end{equation}

However, in selective context-aware sensing, data is collected only when required. This means that the mobile sensing platform does not push data to the cloud all the time (i.e. $T_{TD} $). As shown in Equation \ref{equation:010}, actual running time (i.e. $T_{ART}$) is less than the total duration of the scenario (i.e. $T_{TD} $) due to $T_{\textbf{$\Pi$}}$. $T_{\textbf{$\Pi$}}$ denotes the time period where mobile sensing platform is not interested to push data to the cloud based on the sensing objectives and instructions provided to it (e.g. through context-aware policies)

\begin{equation}
\label{equation:010}
\downarrow T_{ART}= T_{TD} - T_{\Pi} \uparrow
\end{equation}

As a result, actual running time ($T_{ART}$) is less than total duration of a given scenario ($T_{TD}$) as shown in Equation \ref{equation:011}.

\begin{equation}
\label{equation:011}
T_{ART} < T_{TD}
\end{equation}

Then, we can define the energy consumption related to data communication ($E_{DCom}^{S^{\Psi}}$)  for a given scenario $\Psi$ which employs context-aware capabilities to reduce energy wastage as in Equation \ref{equation:013} and \ref{equation:012}.

\begin{equation}
\label{equation:013}
TT_{DCom}^{S^{\Psi}}  = \frac{T_{ART}^{S^{\Psi}} }{T_{NCF}^{S^{\Psi}}  } \times T_{DCom}^{\alpha }
\end{equation}

\begin{equation}
\label{equation:012}
E_{DCom}^{S^{\Psi}} =  E_{DCom}^{\Delta} \times TT_{DCom}^{S^{\Psi}}
\end{equation}

Finally, by applying Equation \ref{equation:012} to Equation \ref{equation:03} and Equation \ref{equation:02}, we can model the total saving as defined in Equation \ref{equation:014}.

\begin{equation}
\label{equation:014}
\textrm{Total Energy Cost Saving}  = \frac{E_{Total}^{S^{\Psi}}}{E_{Total}^{S^{\Theta}} } 
\end{equation}

\subsection{Storage Consumption Modelling}
\label{sec:Storage}

In a similar  way to energy consumption modelling, we can also model storage consumption. In a non-context-aware $\Theta$ scenario, the total number of sensor data records collected are denoted by $N^{\Theta}_{Records}$. Storage frequency is denoted by $T_{SF}^{S^{\Theta}}$ and the total duration of the scenario $\Theta$  is denoted by $T_{TD}^{S^{\Theta}}$. Therefore, Equation \ref{equation:016} defines the number of records that will be stored during the scenario $\Theta$.

\begin{equation}
\label{equation:016}
N^{\Theta}_{Records} = \frac{T_{TD}^{S^{\Theta}} }{T_{SF}^{S^{\Theta}}  }
\end{equation}

The total storage requirement is defined in Equation \ref{equation:017}. It can be calculated by multiplying the  number of records need to be stored and the storage requirement of a single record (i.e. $S_{Record}^{\alpha}$).

\begin{equation}
\label{equation:017}
S_{Total}^{S^{\Theta}} = N^{\Theta}_{Records} \times S_{Record}^{\alpha }
\end{equation}

In Equation \ref{equation:010}, we showed the reduction of actual running time of a context-aware scenario (i.e. $\Psi$). This means that the mobile sensing platform now needs to store  less number of records compared to a $\Theta$ scenario as defined in Equation \ref{equation:018}.

\begin{equation}
\label{equation:018}
N^{\Psi}_{Records} = \frac{T_{ART}^{S^{\Psi}} }{T_{SF}^{S^{\Psi}}  }
\end{equation}

As a results, less number of records require less amount of storage as denoted in Equation \ref{equation:019}. Finally, we can model the storage cost savings  as  in Equation \ref{equation:020}.

\begin{equation}
\label{equation:019}
S_{Total}^{S^{\Psi}} = N^{\Psi}_{Records}  \times S_{Record}^{\alpha }
\end{equation}

\begin{equation}
\label{equation:020}
\textrm{Total Storage Cost Saving}  = \frac{S_{Total}^{S^{\Psi}}}{S_{Total}^{S^{\Theta}}} 
\end{equation}

\subsection{Network Communication Modelling}
\label{sec:Network_Communication}

In the above section, we showed how context-aware $\Psi$ scenarios can reduce storage consumption. Based on that, we base our argument that network communication mostly has the same characteristics as storage. This means that more data we save, it costs more to transfer them to the IoT cloud. Based on Equations \ref{equation:021}, \ref{equation:022}, \ref{equation:023}, \ref{equation:024}, we can infer, context aware capabilities lead to reduce network communication wastage.

\begin{equation}
\label{equation:021}
NC_{Total}^{S^{\Theta}}   \approx S_{Total}^{S^{\Theta}}
\end{equation}

\begin{equation}
\label{equation:022}
NC_{Total}^{S^{\Psi}}  \approx S_{Total}^{S^{\Psi}}
\end{equation}

\begin{equation}
\label{equation:023}
S_{Total}^{S^{\Psi}}  < S_{Total}^{S^{\Theta}}
\end{equation}

\begin{equation}
\label{equation:024}
NC_{Total}^{S^{\Psi}} < NC_{Total}^{S^{\Theta}} 
\end{equation}

Finally, Equation \ref{equation:025} models the total network communication savings of C-MOSDEN when employing context-aware capabilities. 

\begin{equation}
\label{equation:025}
\textrm{Total Network Communication Saving}  = \frac{NC_{Total}^{S^{\Psi}}}{NC_{Total}^{S^{\Theta}} } 
\end{equation}

In above three sections, we theoretically explained how and why context-aware selective sensing is more efficient over non-selective non-context aware sensing strategies. In Section \ref{sec:Implementation} and \ref{sec:Evaluation}, we validate the theoretical modelling by conducting a series of experimental evaluations.

\begin{figure*}[t!]
        \centering
        \begin{subfigure}[b]{165pt}
                \centering
                \includegraphics[scale=0.38]{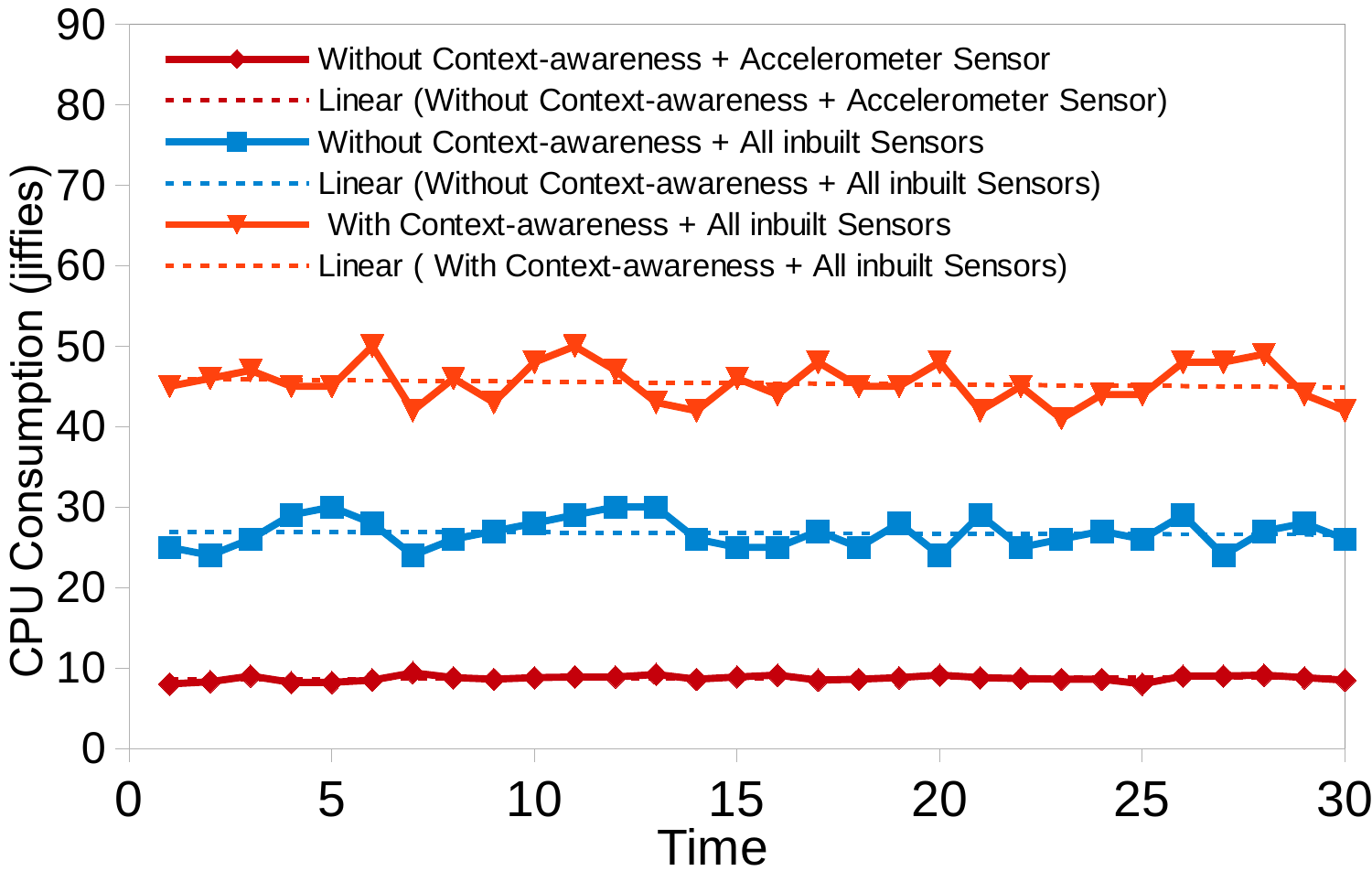}
                \caption{\footnotesize }
                \label{Figure:Overhead_CPU}
        \end{subfigure}%
        ~ 
        \begin{subfigure}[b]{165pt}
                \centering
                \includegraphics[scale=.38]{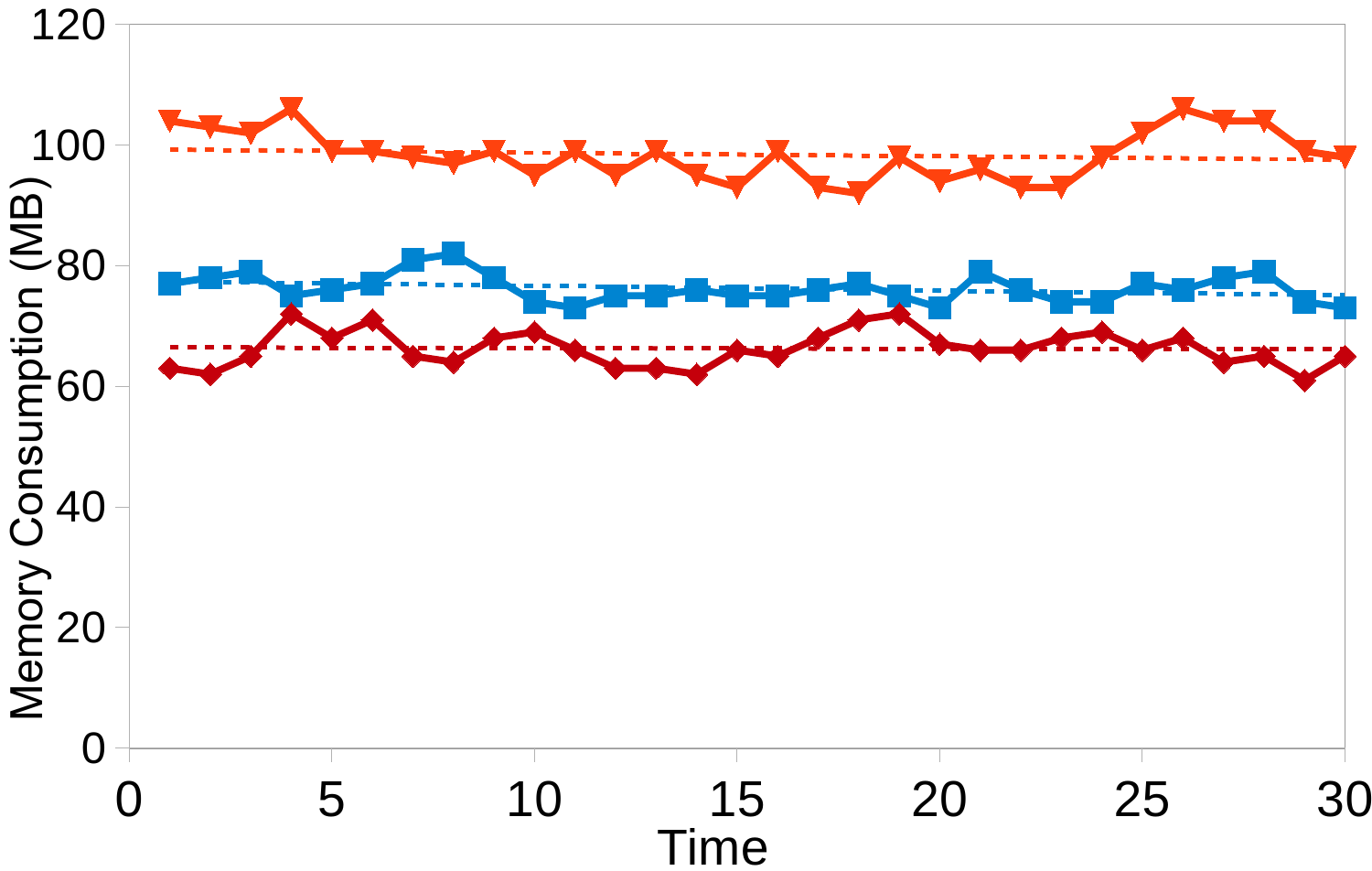}
                \caption{\footnotesize }
                \label{Figure:Overhead_Memory}
        \end{subfigure}
        ~ 
        \begin{subfigure}[b]{165pt}
                \centering
                \includegraphics[scale=.38]{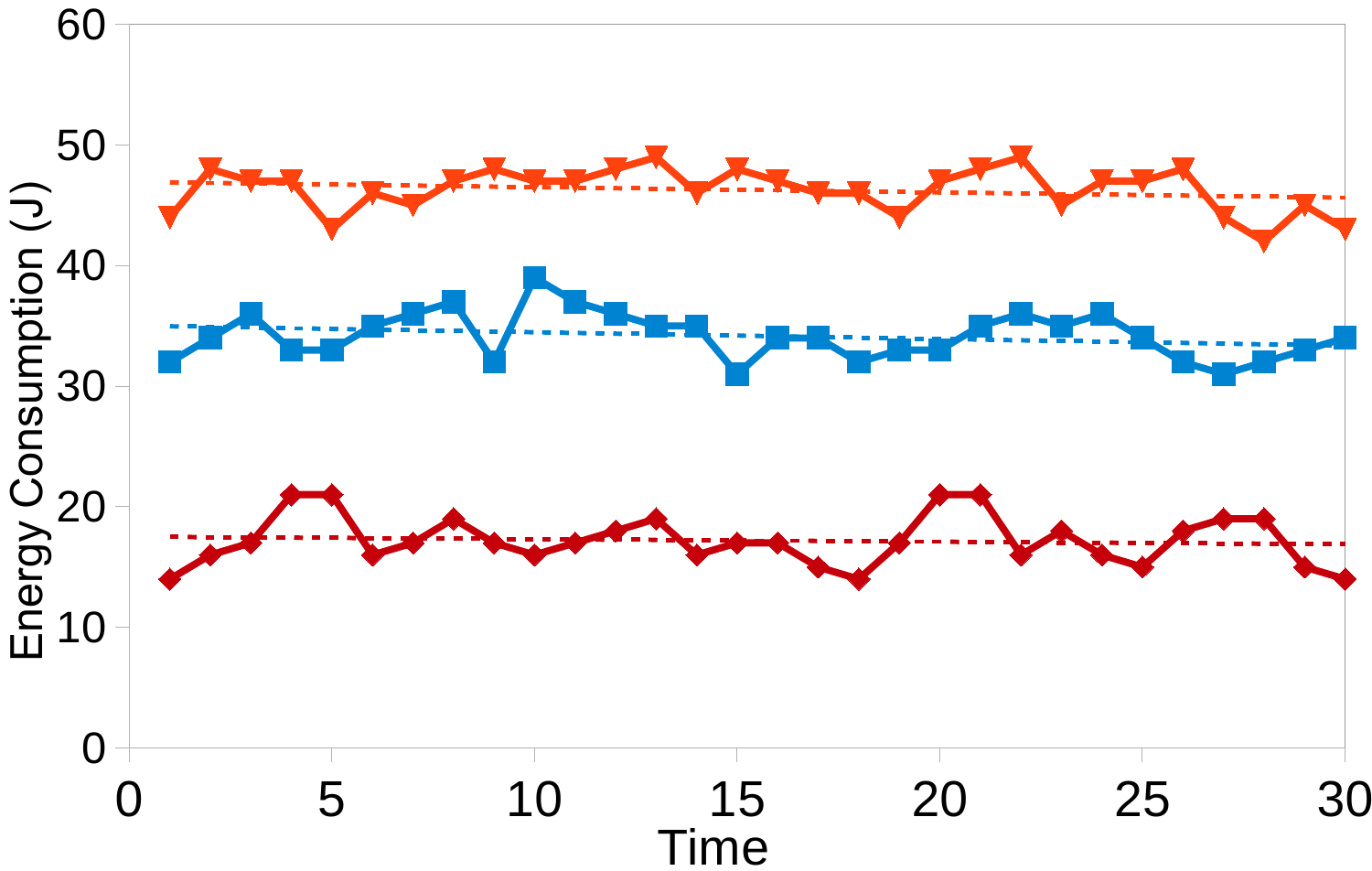}
                \caption{\footnotesize }
                \label{Figure:Overhead_Energy}
                
        \end{subfigure}
        ~ 
          
         \begin{subfigure}[b]{165pt}
         \centering
         \includegraphics[scale=0.38]{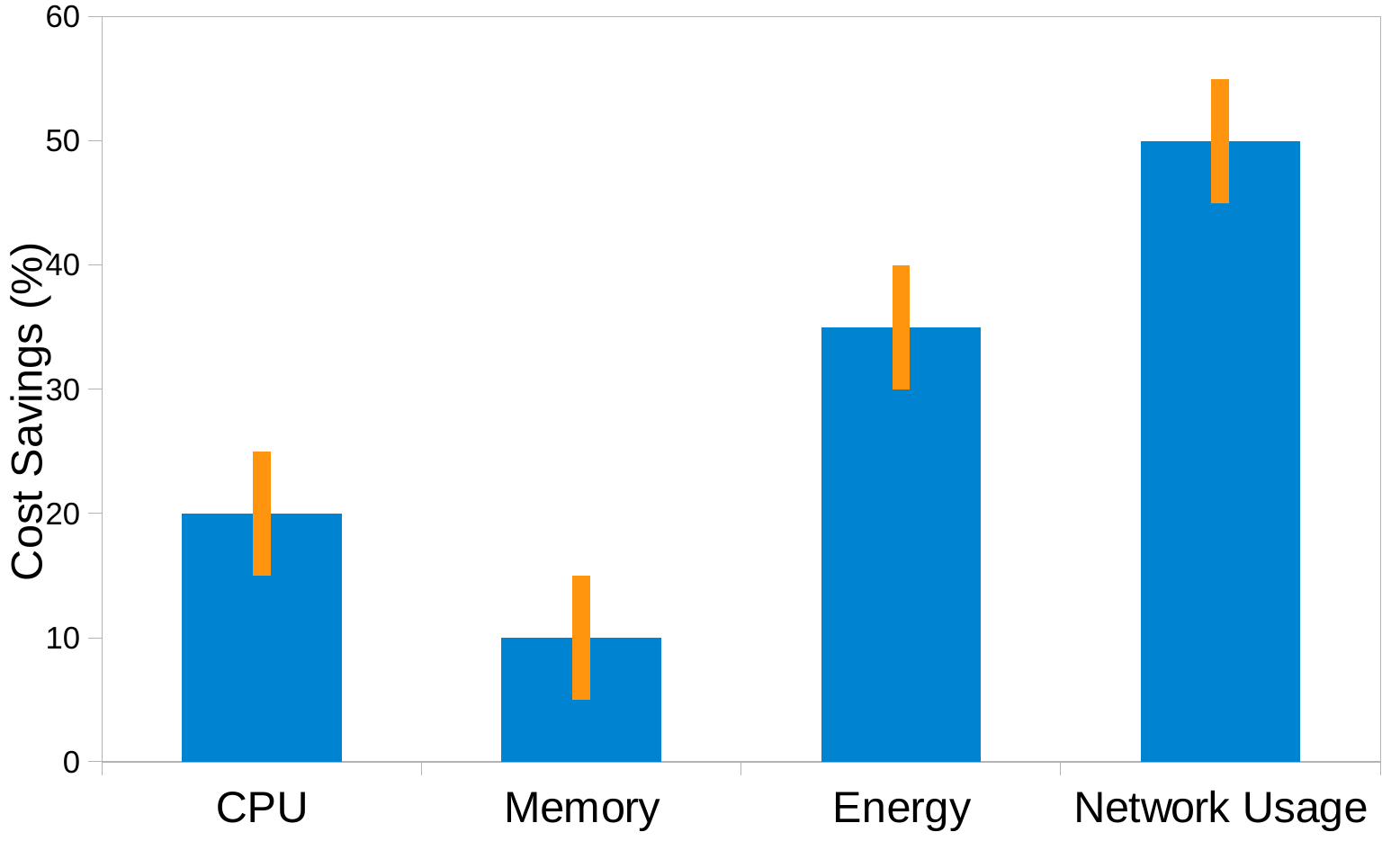}
          \caption{\footnotesize }
          \label{Figure:Cost_Saving_Real}
         \end{subfigure}%
        \begin{subfigure}[b]{165pt}
                \centering
                \includegraphics[scale=.38]{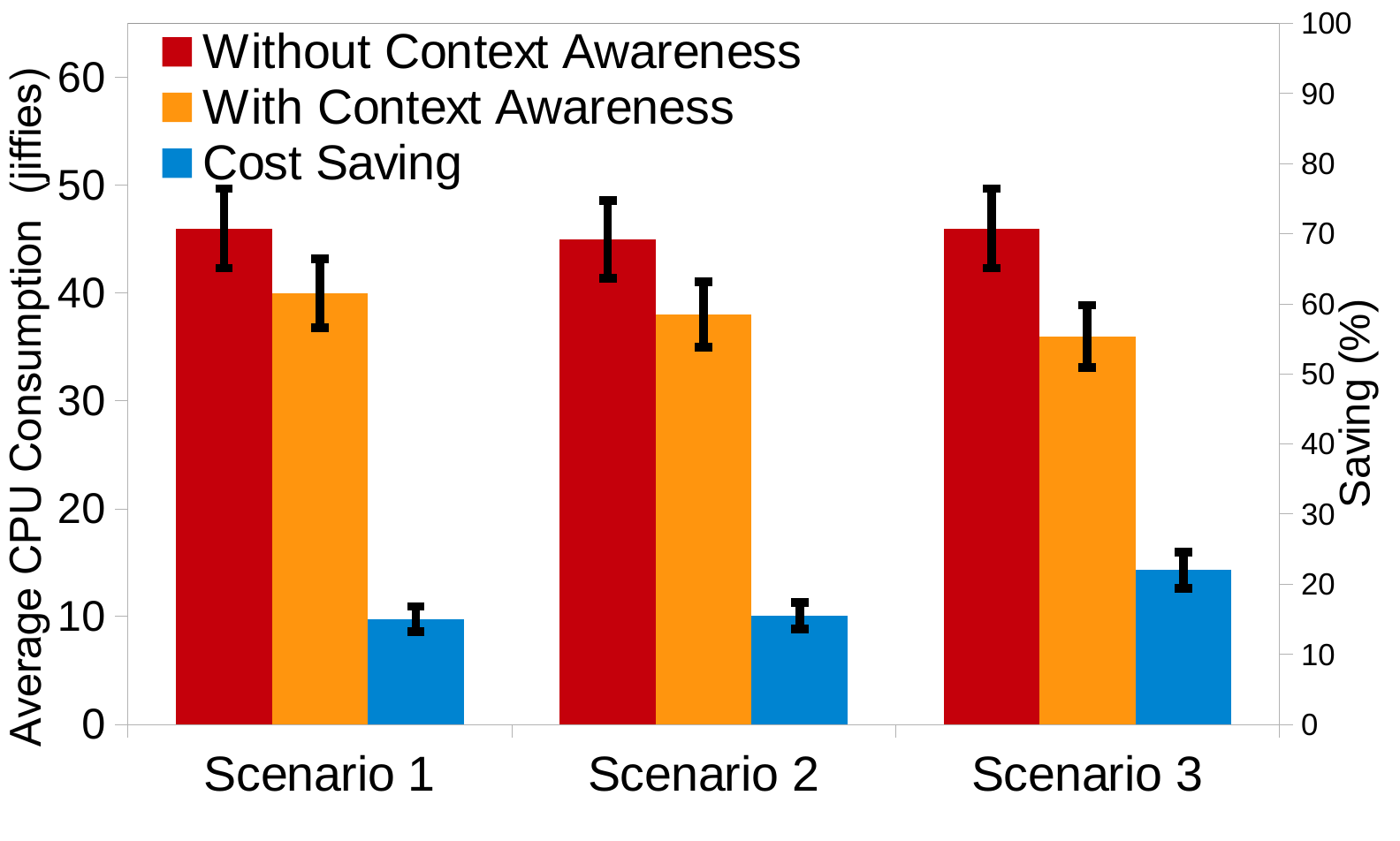}
                \caption{\footnotesize }
                \label{Figure:Cost_Saving_CPU}
        \end{subfigure}
        ~ 
        \begin{subfigure}[b]{165pt}
                \centering
                \includegraphics[scale=.38]{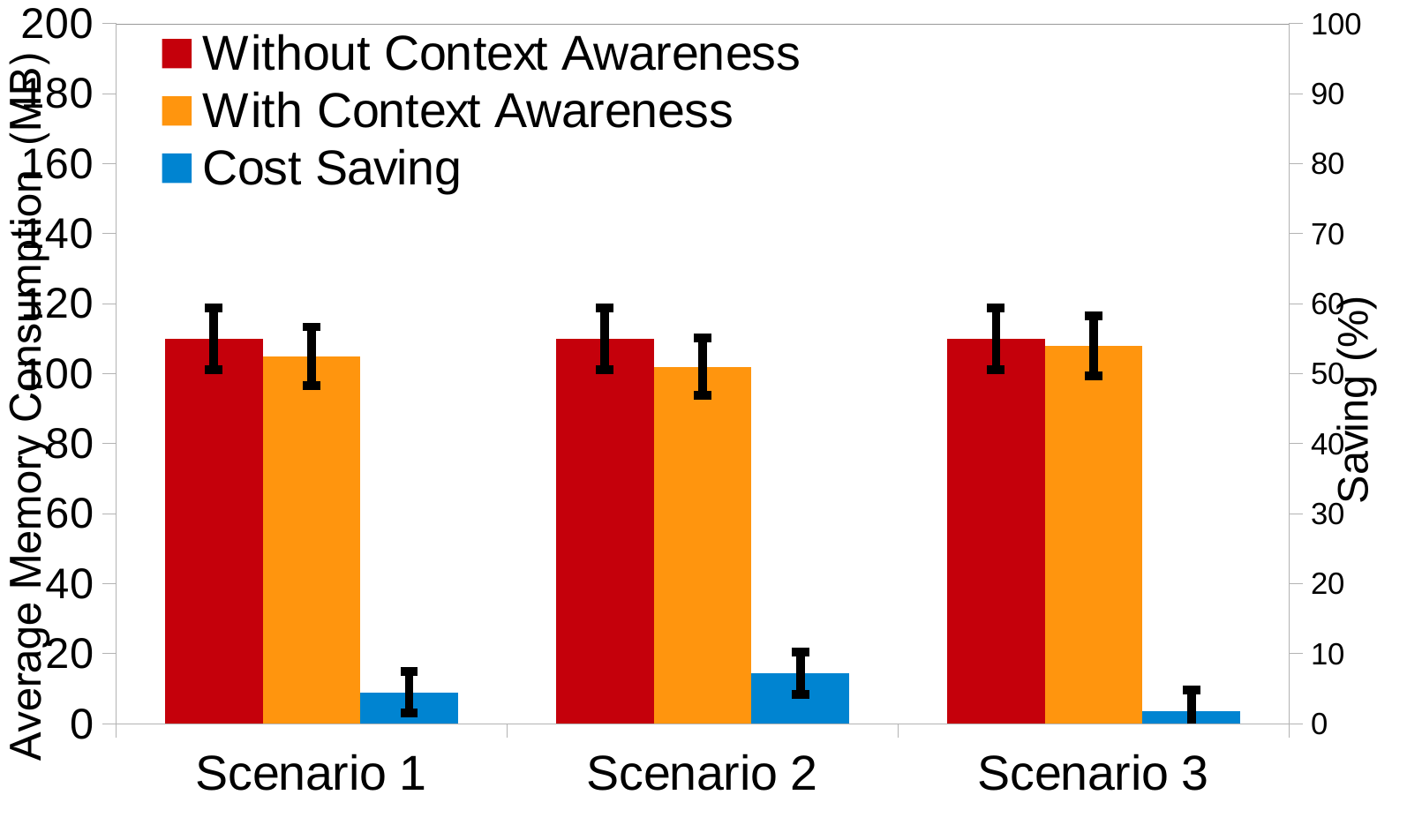}
                \caption{\footnotesize }
                \label{Figure:Cost_Saving_Memory}
        \end{subfigure}

                ~ 
                  
             \begin{subfigure}[b]{165pt}
                \centering
                \includegraphics[scale=.38]{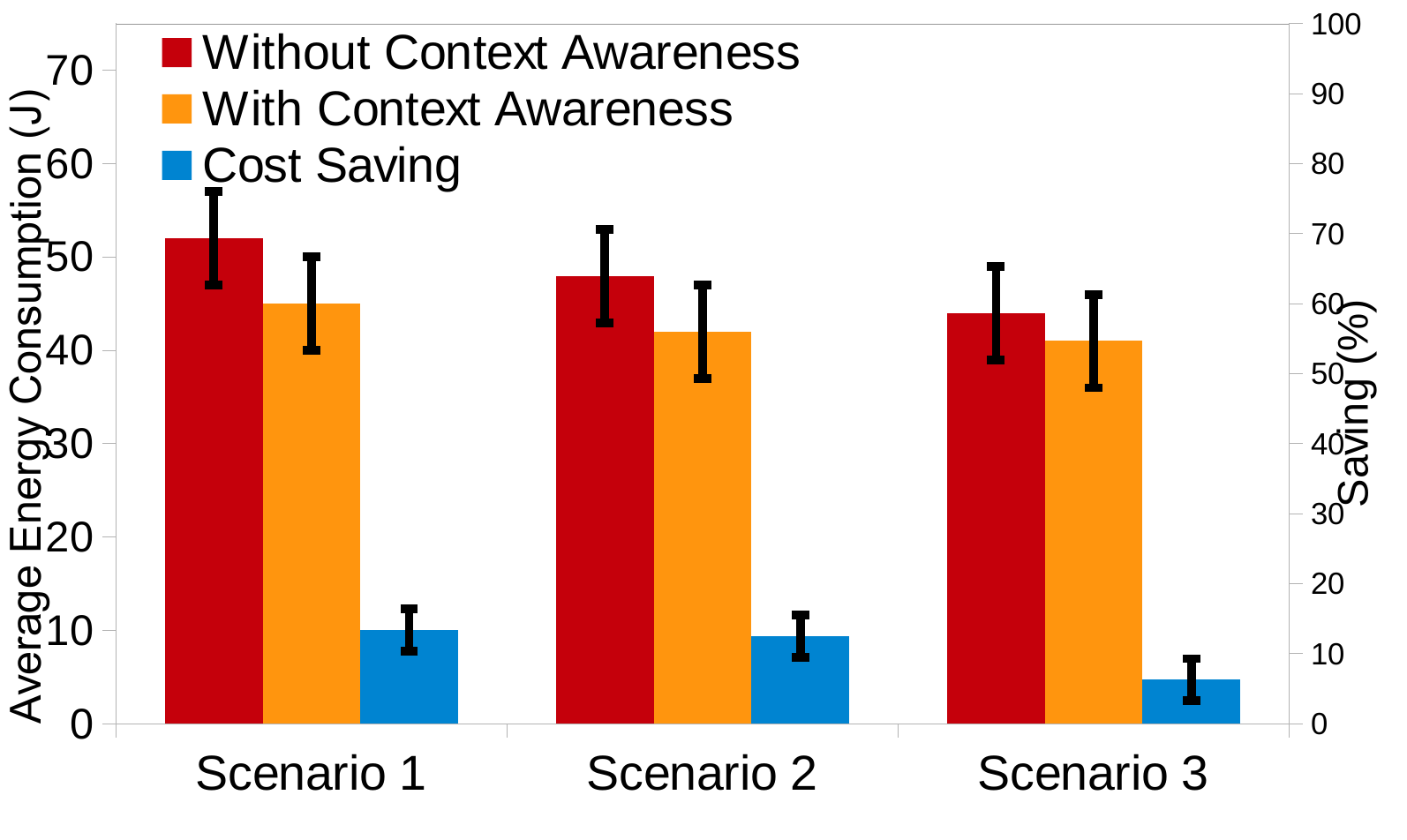}
                \caption{\footnotesize }
                \label{Figure:Cost_Saving_Energy}
                          \end{subfigure}  
                \begin{subfigure}[b]{165pt}
                        \centering
                        \includegraphics[scale=.38]{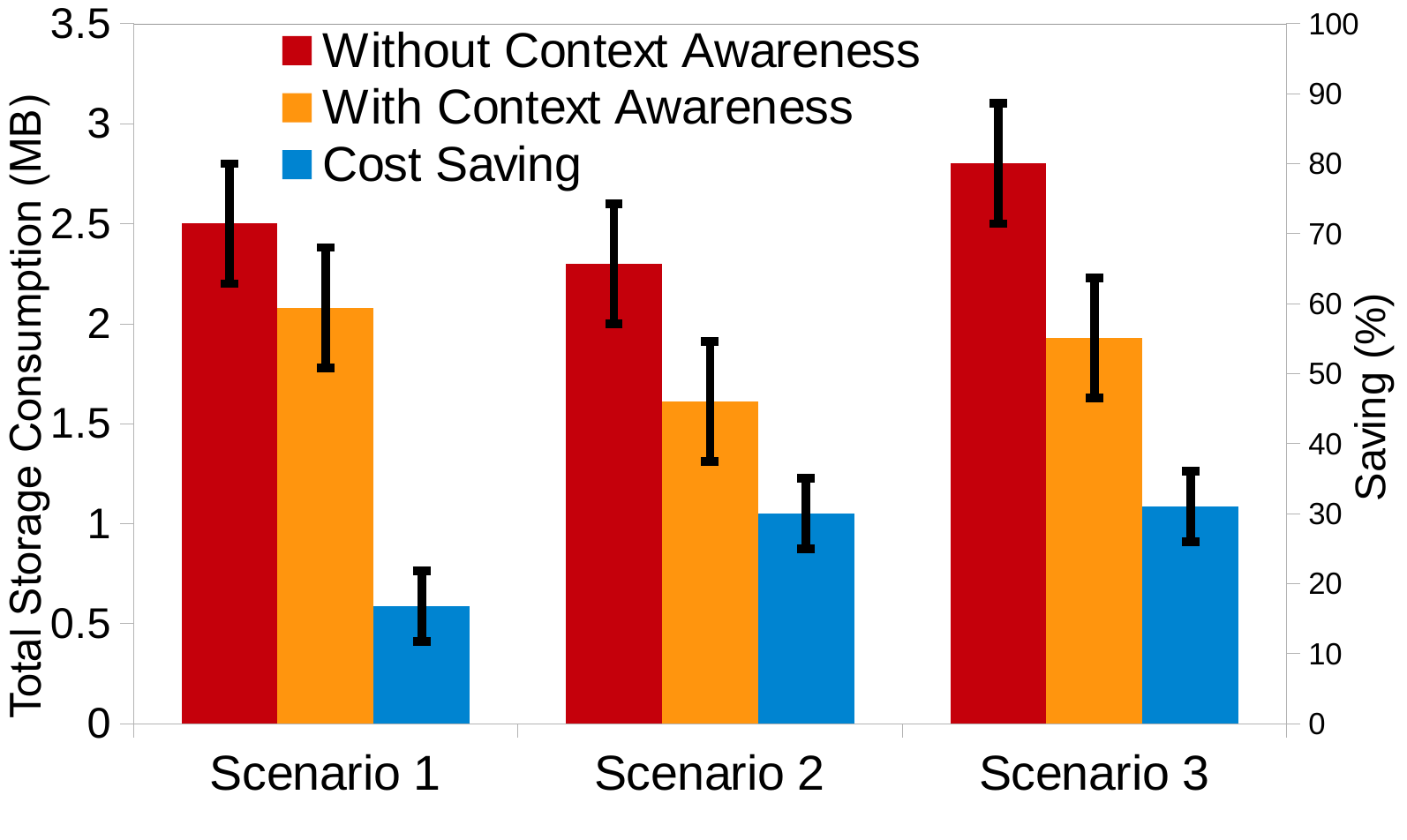}
                        \vspace{-16pt}
                        \caption{\footnotesize }
                        \label{Figure:Cost_Saving_Storage}
                \end{subfigure}
                ~ 
                \begin{subfigure}[b]{165pt}
                        \centering
                        \includegraphics[scale=.38]{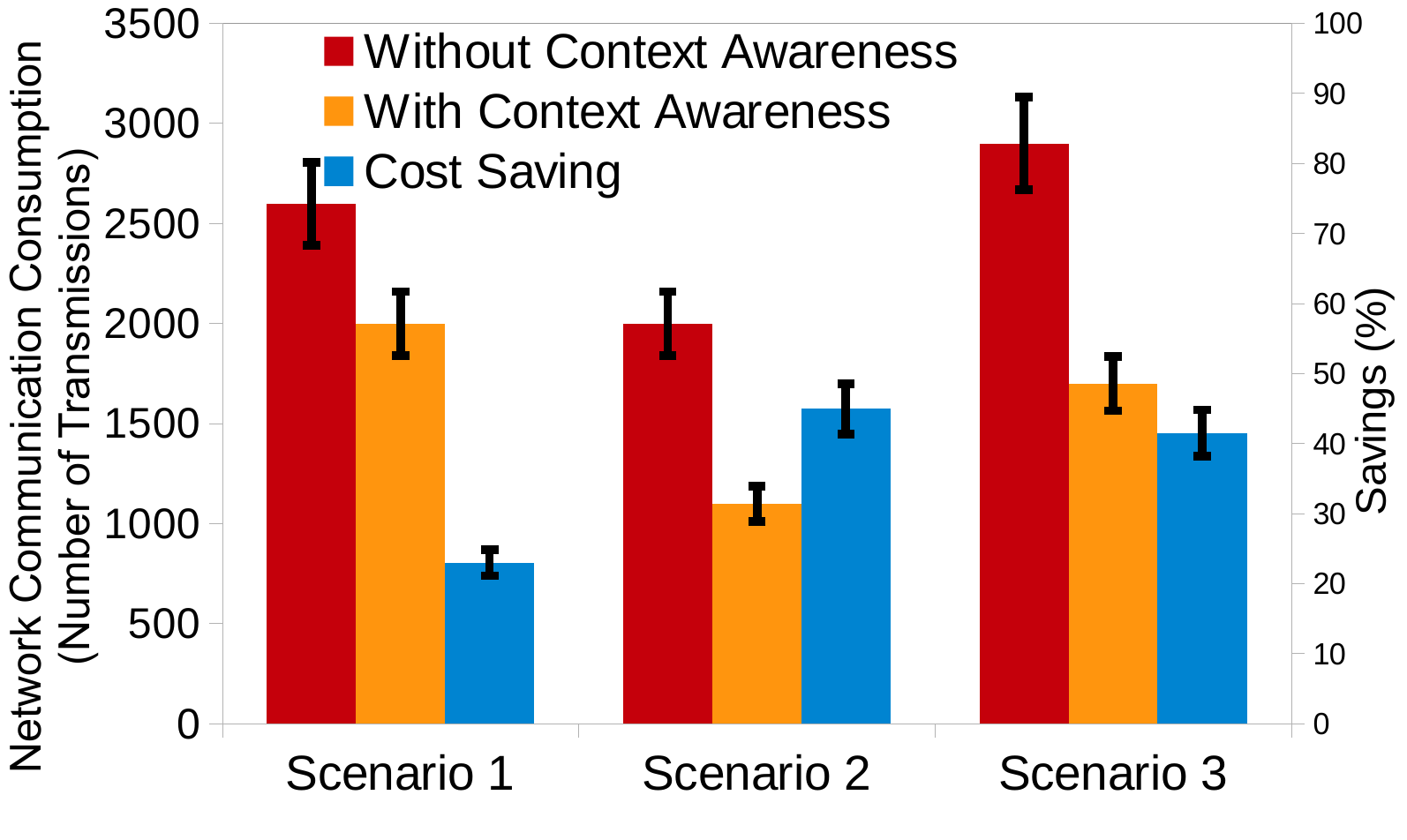}
                        \vspace{-16pt}
                        \caption{\footnotesize }
                        \label{Figure:Cost_Saving_Network_Communication}
                \end{subfigure}
                \label{fig:animals}
        \caption{C-MOSDEN Performance Evaluation}
        \vspace{-20pt}
\end{figure*}

\section{Implementation and Experimental Test-bed}
\label{sec:Implementation}

This proposed context-aware mobile sensing platform, C-MOSDEN, is developed using the Android platform. We used a Google Nexus 4 device with the Android KitKat operating system for evaluations. Hardware  of the device consists of Qualcomm Snapdragon S4 Pro CPU, 2 GB RAM and 16GB storage. In order to verify the hypothesis and the cost models, we ran different usecase experiments as explained below in Section \ref{sec:Evaluation}. Each usecase is run 10 times and average has been presented as the results. We employed a human actor to perform the three scenario mentioned in the Section \ref{sec:Functional}. More importantly, we also ran  each usecase under different configurations (e.g. different number of sensors used  depending on the experiment) to create a benchmark for comparison. We also ran the experiment with and without context-aware capabilities. We built the current activity-aware module using the Android SDK  API. We also created a location-aware module by using geofencing technique provided in Google API\footnote{\url{https://developer.android.com/reference/com/google/android/gms/location/package-summary.html}}. It is important to highlight  that our intention is not to develop a more efficient activity  recognition or geofencing technique, but to use activity and location awareness to enable selective-sensing in sensing as a service domain towards improving efficiency. Furthermore, it is very easy to replace the context-aware module that will be employing at a given time. Thus, regardless of the underlying context-aware modules used, the system will always behave the same way from the user perspective. We used a location mocking tool called \textit{'Fake GPS location'} to test geofencing capabilities. In all the evaluations, CPU usage (consumption) is measured in units of jiffies\footnote{In computing, a jiffy is the duration of one tick of the system timer interrupt. It is not an absolute time interval unit, since its duration depends on the clock interrupt frequency of the particular hardware platform}.

\section{Lessons Learned and Discussion}
\label{sec:Evaluation}

In this section, we present details of the  experimental evaluations that are  performed using the proposed sensing platform, C-MOSDEN. Our experiments consists of both real-world experiments and simulated lab based experiments.

In the first three series of experiments, our intention was to understand the impact of context-aware functionality towards CPU consumption, memory consumption, and energy consumption. As we highlighted in the cost models, presented in Section \ref{sec:Model}, there are overheads created by activity-aware and location-aware capabilities. 

First, we measured the above mentioned parameters while context-aware capabilities are deactivated and one the accelerometer sensor is configured to collect data. Secondly, we activated all the sensors available in the smart phone (i.e. accelerometer, gravity, gyroscope, liner acceleration, rotation vector,  light, pressure,  magnetic fields, orientation, proximity). We also kept the  context-aware capabilities deactivated. Thirdly, we activated the context-aware capabilities and kept the number of sensors used to collect data the same as before. In this case, we did activated the context-aware capabilities however did not use the capabilities in action. We configured the mobile device to push all the sensed data to the IoT cloud without any context-aware filtering. By doing so, we were aimed to compare the computational requirements with and without context-aware capabilities activated to identify the overhead created by the context-aware reasoning modules.

In these experiments, our objective is to understand how much more computational resources are required by C-MOSDEN when context-aware capabilities are in action. Specially, we changed the number of sensors used to collect data in two different experiments, in order to compare the resource consumption variability when different number of sensors are activated, in comparison to when the context-aware capabilities are activated. It is also important to note that, in these experiments, we configured C-MOSDEN to collect sensor data and push to the IoT cloud middleware in one second intervals. We ran the experiments for 30 minutes. The results presented in  Figure \ref{Figure:Overhead_CPU} (CPU consumption), Figure \ref{Figure:Overhead_Memory} (memory consumption), and Figure \ref{Figure:Overhead_Energy} (energy consumption).

\textcolor{black}{According to the results, it is evident that context-aware functionalities creates some overhead in term of CPU, memory and energy. Based on our experience, in this paper as well as in the past \cite{ZMC008}, CPU and energy consumptions are not very good indicators to measure the computational complexity, specially in Android, due its auto load balancing of computational requirement between different applications. However, memory is  a much better  indicator to measure the computational complexity. Android allocates memory less greedily to application as long as it has abundant amount of memory which is 2GB in Nexus 4 device. If we analyse the memory consumption results in Figure \ref{Figure:Overhead_Memory}, it is evident that, additional overhead created by context-aware functionalities are not substantial.}

Next, we ran an experiment to test the capabilities of C-MOSDEN in the real world. In this experiment, we measured CPU, memory, and energy consumptions, and network usage. In order to plot all the result in a single graph, we used the cost savings as the common measurements  (as a percentage). The results are presented in Figure \ref{Figure:Cost_Saving_Real}.  In this experiments. First, the user walked for 10 minutes. Then, he cycled for 20 minutes and then  walked again for another 10 minutes. We ran the experiment with both while context-aware capabilities were ON and OFF. Mobile phone has been configured to collect sensor data using all available sensors. The objective was to collect sensor data only when the users is cycling.

\textcolor{black}{According to the results presented in Figure \ref{Figure:Cost_Saving_Real}, it is evident that context-aware capabilities have been able to save costs in term of all four parameters we measured. However, the most significant saving is energy and network usage. Both energy and network usage have been significantly reduced due to selective sensing. Energy consumption is reduced due to the less usage of wireless communication radios \cite{TCSS3}.}

It is important to note that running  these experiments required significant amount of time and budget. Therefore, we decided to simulate the scenario we presented earlier in this paper in a lab environment. However, the result we gathered in this real-world example substantially validates our lab simulations.
Later, we ran series of experiments to evaluate three usecase scenarios presented in Section \ref{sec:Functional}. We simulated those scenarios in a lab environment. The data collecting specifications are as follows. All the available sensors were configured to collect data in following experiments. We conducted the experiment both with and without context-aware capabilities activated. We measured CPU, memory, energy, storage, and network consumption. Results are presented  in Figures \ref{Figure:Cost_Saving_Real} to Figure \ref{Figure:Cost_Saving_Network_Communication} respectively. To run these experiments, we created predefined data set that simulates the relevant use behaviour including location changes and activities changes over time.

\begin{itemize}
\item Scenario 1 (Environmental Monitoring):Bus moves 5 minutes and stops for 2 minutes. This pattern will continue for 60 minutes (1.e. 10 stops). The sensing objective is to collect sensor data only when bus is moving. The total duration of the experiment 6o minutes.

\item Scenario 2 (Rehabilitation): The patient performs medically recommended walking exercises for 20 minutes and rest for 15 minute. Then, the patient again walks for 15 minutes. The sensing objective is to collect sensor data only when the patient is walking. The total duration of the experiment 50 minutes.

\item Scenario 3 (Health and Well-being): The user cycles to the jogging path for 10 minutes and then she jogs for 30 minutes. Next she does some bar exercise for 15 minutes before return home by cycling (another 10 minutes). The sensing objective is to collect sensor data only when uses is jogging in the jogging path. The total duration of the experiment is 65 minutes.
\end{itemize}

\textcolor{black}{According to the results presented in Figure \ref{Figure:Cost_Saving_CPU} to \ref{Figure:Cost_Saving_Network_Communication}, it is evident that context-aware capabilities can save costs at different levels depending on the scenario, sensing objectives, conditions, and characteristics. Based on the results gathered in these experiments, we can  conclude that any kind of context-aware functionalities (e.g. time-awareness and social awareness) that would reduce the uninterested data collection and transmission can be helpful to save costs. }

\textcolor{black}{In general, wireless communication radios switching on and off consumes significant amount of energy. If the number of times these radios switched on can be reduced, it helps to significantly  reduce the energy consumptions. As shown in theoretical models, lesser the amount of data is captured, the less time it will take to transfer the data over the cloud, so the communication radios will only be required for shorter durations. When wireless radios are not actively transmitting data, they will also put less workload on the CPU as well due to less reads/writes from the storage (which also requires less memory). By conducting a number of experiments, we have comprehensively validated the theoretical models presented in Section \ref{sec:Functional}. We have also verified the importance of context-aware capabilities integrated into mobile sensing platforms in order to breakdown Big Data into small data so anyone can analyse them and derive knowledge  easily with less amount of resources and budgets.}


\section{Related Work}
\label{sec:Related_Work}

 Mobile phone based sensing algorithms, approaches, and applications are discussed in \cite{P217}. DAM4GSN \cite{ZMP001} is  an approach based on GSN that is capable of collecting data from internal sensors of a mobile phone and sending it to the GSN middleware. No processing capabilities are provided at the mobile phone end. Therefore, all the information sensed is sent to the server. This approach is inefficient due to the continuous usage of the communication radio of the mobile phone and may also communicate sensor data that are not required or important to the data sensor data consumer \cite{ZMP001}. \textit{Dynamix} \cite{P627} is a plug-and-play context framework for Android. \textit{Dynamix} automatically discovers, downloads and installs the plug-ins needed for a given context sensing task. \textit{Dynamix} is a stand alone application and it tries to understand new environments by using pluggable context discovery and reasoning mechanisms.  Context discovery is the main functionality in \textit{Dynamix}.

One of the most popular type of processing in mobile is activity recognition. Yan et al. \cite{P629} have presented an energy-efficient continuous activity recognition on mobile phones. Choudhury et al. \cite{Z1016} has also developed customs mobile sensing hardware platform for activity recognition. Activities such as walking, running taking stairs up/down, taking elevator up/down, cooking, working on computer, eating, watching TV, talking, cycling, using an elliptical trainer, and using a stair machine can be detected by using the device. Choudhury et al. have used sensors such as microphone, light, 3-axis digital accelerometer, barometer temperature, IR and visible+IR light, humidity/temperature, Compass,  3D magnetometers, 3D gyroscope, and 3D compass to collect data to support their algorithms that detect the activities. Lee et al. \cite{Z1017} have developed a similarity system. However, instead of processing the data in the mobile device, it sends data to the cloud by using a smartphone as an intermediate gateway device. Another similar approach has been presented by Laukkarinen et al. \cite{Z1018}. They have implemented a distributed middleware for 8-bit micro controller nodes where executing instructions (e.g. for data processing and event detection) are sent to each node using a \textit{Process Description Language} (PDL). It is important to note that all these approaches focus on building activity recognition modules. In contrast, we employ an activity recognition module to filter unnecessary data processing and communication with the intention of reducing all costs. CONSORTS-S \cite{Z1020} has also used a similar approach. Instead of getting data from external sensors directly into mobile phones, CONSORTS-S uses a custom made sensor board that connect to the mobile phone using a serial cable which allows the mobile phone to collect data from external sensors.

Most mobile sensing applications can be classified into \textit{personal} and \textit{community sensing} \cite{P217}. \textit{Personal sensing} applications focus on the individuals. On the contrary, \textit{community sensing} also termed \textit{opportunistic/crowdsensing}\footnote{In this chapter, we use the terms \textit{opportunistic sensing }, \textit{crowdsensing } and \textit{participatory sensing }synonymously.} takes advantage of a population of individuals to measure large-scale phenomenon that cannot be measured using single individual. In most cases, the population of individuals participating in \crowdsensing applications share a common goal. To date, most efforts to develop \crowdsensing applications have focused on building monolithic mobile applications that are built for specific requirements \cite{pogo}. Furthermore, the sensed data generated by the application are often available only within the closed population \cite{P485}. However, to realise the greater vision of a collaborative mobile \crowdsensing application, we would need a common platform that facilitates easy development and deployment of collaborative crowd-sensed applications \cite{Z1034}.

Grid-M \cite{Z1019} is a platform for lightweight grid computing. It is a tailored for embedded and mobile computing devices. The middleware is built using  Java 2 Micro Edition, and an application programming interface (API) is provided to connect Java-developed applications in a Grid Computing environment. This work highlights the importance of providing and API based communication channel which enables communication. As illustrated in Figure \ref{Figure:Sensing_as_a_service}, mobile nodes work  similar to grid computing,  where they work together to collect sensors data as instructed by the cloud based IoT middleware or by their own peers (e.g. other mobile sensing platform nodes). Zhang et al. \cite{Z1021} have developed a middleware on top of TinyOS (tinyos.net)  for TelosB sensors. The data fusion components are designed as agents which they migrate form one node to another. Such migration is an efficient technique in term of resource utilization. Data fusion consumes the resources only when a given node required to process data. Otherwise the agents moves on to another node on demand. We simulate such behaviour in C-MOSDEN where plugins are installed when needed and uninstall when not needed. Another agent-based sensing platform has been proposed by Sun and Nakata \cite{Z1026}. Budde et al. \cite{Z1030} have proposed a framework that allows to discover  smart objects in the Internet of Things. The framework allows smart objects and services to be registered by providing metadata where it later allows searching and selection. Mori et al. \cite{Mori} has proposed a cloud-based mobile phone sensing middleware \cite{TCSS2} that can collectively sense the environment as group of participants. however, if there are more participants present in a given region that expected, the task will be selectively assigned to the most appropriate participants by considering context information such as remaining energy, exact location, and so on. Their approach is also focusing on reducing unnecessary amount of data capturing and communication.

NORS \cite{Z1023} is an open source platform that enables participatory sensing using mobile phones. It mainly focuses on collecting data instead of processing. The platform includes external sensors, mobile phones, and a cloud service for data storage. Sharing data among of mobile phones is not supported. In contract, C-MOSDEN is capable of peer to peer communication as well as cloud based communication. USense \cite{Z1024} is client-side middleware that opportunistically and passively (i.e. without human intervention) performance sensing tasks in crowd sensing fashion. It uses XML definitions to explain a \textit{`moment'} where the middleware needs to start sensing and stop sensing. The \textit{`moment'} are composed with a bunch of condition such as location, time, and so on. Similarly, SENSE-SATION \cite{Z1025} also gathers and stores sensor information using mobile phones and make them directly accessible over the Internet via RESTful web services. Patti et al. \cite{Patti} have proposed an energy-efficient middleware aims at improving energy efficiency of public buildings and spaces exploiting both event-driven and user centric approaches. In their work, sensors are used to detect user presence. Then, system actuates heating systems accordingly to reduce energy wastage.

\section{Conclusions and Future Work}
\label{sec:Conclusions and Future Work}

We have presented our \platform platform to support on-demand distributed mobile crowd sensing. Our objective was to built a platform that can perform sensing tasks in a collaborative and selective manner. For example, the \platform platform can be remotely configured to sense only when a certain activity occurs (e.g. driving, running, walking).  Further, the C-MOSDEN platform supports location-aware sensing (e.g. sense only when a user enters to a particular building). Moreover, the platform  has the capability to autonomously select which communication channel (e.g. WiFi or 3G) to use to send the data to the cloud based on context information such as battery level and availability. The proposed platform collects only the data that are relevant to the data consumers, thereby reducing the data storage requirements and processing requirements. We discussed three different real-world use case scenarios where the proposed platform can offer significant advantages. It was shown to  facilitate the efficient and effective mobile crowd sensing functionality at a  minimum cost. Through a series of experimentations and evaluations, we showed the importance of selective sensing through the reduction  of computational requirements. In general, through selective sensing,  we were able to successfully reduce the energy consumption, network communication requirements and storage requirements. Although the context-aware functionalities have generated a small amount of overhead, it was revealed that the  cost savings and benefits   far outweighed the increased complexity. In future works, we are planning to enrich C-MOSDEN with privacy preserving data analytics capabilities.



%

\ifCLASSOPTIONcaptionsoff
  \newpage
\fi



%
%
%


\def\IEEEbibitemsep{0pt plus .5pt}
\bibliography{Bibliography}
\bibliographystyle{IEEEtran}


\vspace{-30pt}

\begin{IEEEbiography}[{\includegraphics[width=1in,height=1.25in,clip,keepaspectratio]{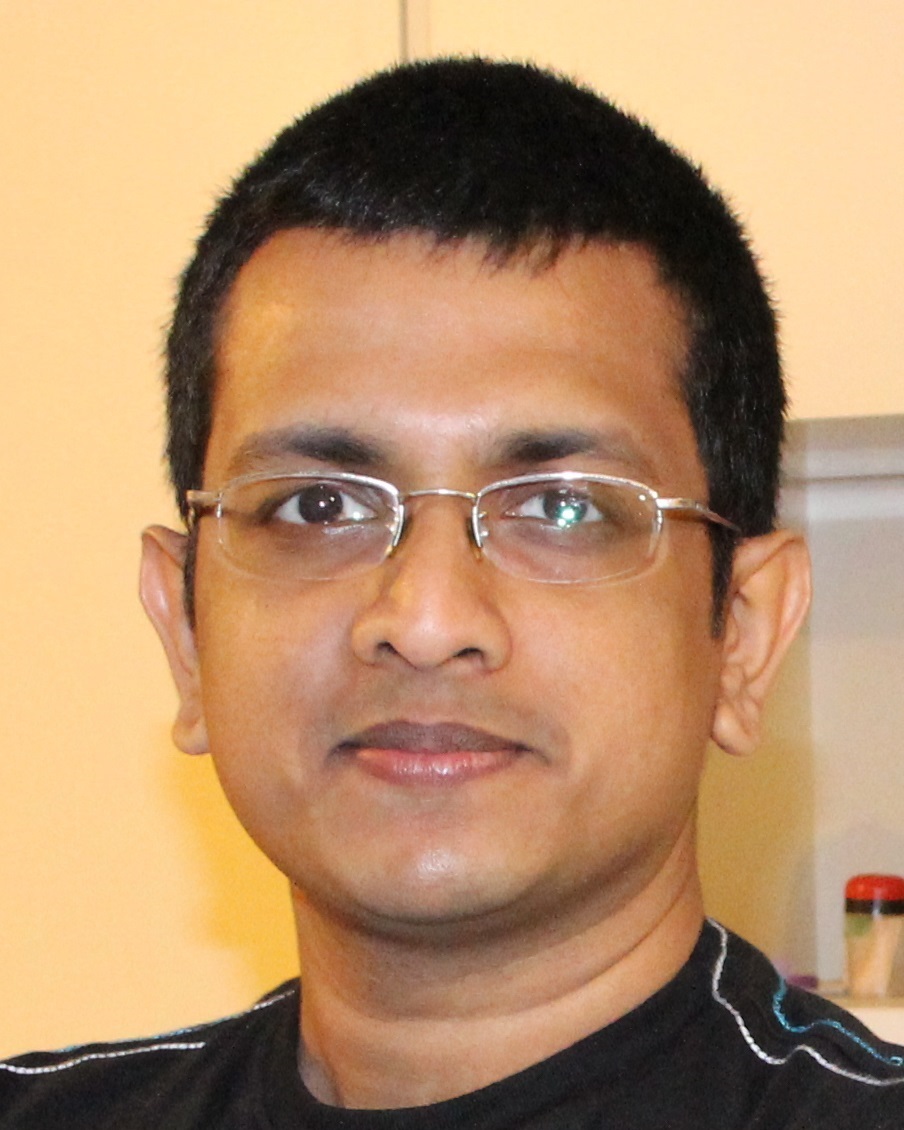}}]{Charith Perera}(S'11-M'14)
received his BSc (Hons) in Computer Science in 2009 from Staffordshire University, Stoke-on-Trent, United Kingdom and MBA in Business Administration in 2012 from University of Wales, Cardiff, United Kingdom and PhD in Computer Science in 2014 from The Australian National University, Canberra, Australia. He is also worked at Information Engineering Laboratory, ICT Centre, CSIRO and involved in OpenIoT Project  which is co-funded by the European Commission under seventh framework program. Currently, he is a Post-doctoral Research Fellow at Open University, UK. His research interests include Internet of Things, Smart Cities, Sensing as a Service, Privacy, Sensing Middleware Architecture. He is a member of both IEEE and ACM.

\end{IEEEbiography}

\vspace{-30pt}

\begin{IEEEbiography}[{\includegraphics[width=1in,height=1.25in,clip,keepaspectratio]{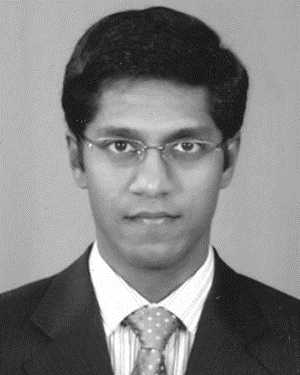}}]{Dumidu S. Talagala}
 (S'11-M'14) received the B.Sc. Eng (Hons) in electronic and telecommunication engineering from the University of Moratuwa, Sri Lanka, in 2007. From 2007 to 2009, he was an Engineer at Dialog Axiata PLC, Sri Lanka. He completed his Ph.D. degree within the Applied Signal Processing Group, College of Engineering and Computer Science, at the Australian National University, Canberra, in 2013. He is currently a research fellow in the Centre for Vision, Speech and Signal Processing at the University of Surrey, United Kingdom. His research interests are in the areas of sound source localization, spatial sound-field reproduction, active noise control, array signal processing and convex optimization.

\end{IEEEbiography}

\vspace{-30pt}

\begin{IEEEbiography}[{\includegraphics[width=1in,height=1.25in,clip,keepaspectratio]{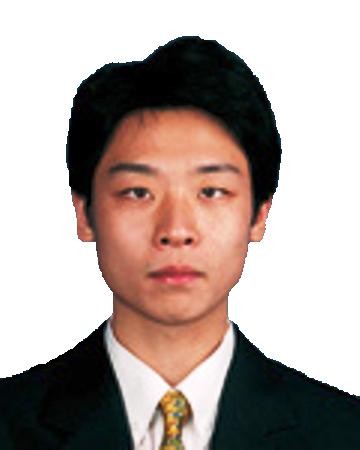}}]{Chi Harold Liu} is a Full Professor at the School of Software, Beijing Institute of Technology, China. He is also the Director of the Institute of Data Intelligence, Director of IBM Mainframe Excellence Center (Beijing), Director of IBM Big Data \& AnalysisTechnology Center, and Director of National Laboratory of Data Intelligence for China Light Industry. He holds a Ph.D. degree from Imperial College, London, U.K., and a B.Eng. degree from Tsinghua University, Beijing, China.  His current research interests include the Internet of Things (IoT), Big Data analytics, and wireless ad hoc, sensor, and mesh networks.  He served as the consultant to Asian Development Bank, Bain \& Company, and KPMG, USA, and the peer reviewer for Qatar National Research Foundation, and National Science Foundation, China. He is a member of IEEE and ACM.

\end{IEEEbiography}

\vspace{-30pt}

\begin{IEEEbiography}[{\includegraphics[width=1in,height=1.25in,clip,keepaspectratio]{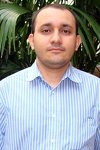}}]{Julio C. Estrella}
 received the Ph.D. in Computer Science at Institute of Mathematics and Computer Science from University of Sao Paulo - USP (2010). MSc in Computer Science at Institute of Mathematics and Computer Science from University of São Paulo - USP (2006). BSc in Computer Science at State University of Sao Paulo -  Julio de Mesquita Filho – UNESP (2002). He has experience in Computer Science with emphasis in Computer Systems Architecture, acting on the following themes: Service Oriented Architectures, Web Services, Performance Evaluation, Distributed Systems, Computer Networks and Computer Security.  He is currently Assistant Professor at Institute of Mathematics and Computer Science - ICMC – USP.

\end{IEEEbiography}
%

%




\end{document}